\documentclass[fleqn,11pt]{SelfArx} 

\usepackage[english]{babel} 

\usepackage{lipsum} 
\usepackage{gensymb}
\usepackage{pdfpages}
\newcommand{\Mearth}{$M_\oplus$}

\DeclareTextSymbol{\deg}{T1}{6}
\DeclareTextSymbol{\deg}{OT1}{23}

\definecolor{color1}{RGB}{0,0,90} 
\definecolor{color2}{RGB}{0,20,20} 

\usepackage{hyperref} 
\hypersetup{hidelinks,colorlinks,breaklinks=true,urlcolor=color2,citecolor=color1,linkcolor=color1,bookmarksopen=false,pdftitle={Title},pdfauthor={Author}}

\JournalInfo{WHITE PAPER RESPONSE TO ESA CALL FOR VOYAGE 2050 SCIENCE THEME} 
\Archive{} 

\PaperTitle{In Situ Exploration of the Giant Planets} 

\Authors{} 

\newcommand{\keywordname}{Keywords} 

\Abstract{Remote sensing observations suffer significant limitations when used to study the bulk atmospheric composition of the giant planets of our solar system. This impacts our knowledge of the formation of these planets and the physics of their atmospheres. A remarkable example of the superiority of in situ probe measurements was illustrated by the exploration of Jupiter, where key measurements such as the determination of the noble gases' abundances and the precise measurement of the helium mixing ratio were only made available through in situ measurements by the Galileo probe. Here we describe the main scientific goals to be addressed by the future in situ exploration of Saturn, Uranus, and Neptune, placing the Galileo probe exploration of Jupiter in a broader context. An atmospheric entry probe targeting the 10-bar level would yield insight into two broad themes: i) the formation history of the giant planets and that of the Solar System, and ii) the processes at play in planetary atmospheres. The probe would descend under parachute to measure composition, structure, and dynamics, with data returned to Earth using a Carrier Relay Spacecraft as a relay station. An atmospheric probe could represent a significant ESA contribution to a future NASA New Frontiers or flagship mission to be launched toward Saturn, Uranus, and/or Neptune.}

\begin{document}
\includepdf[pages=-]{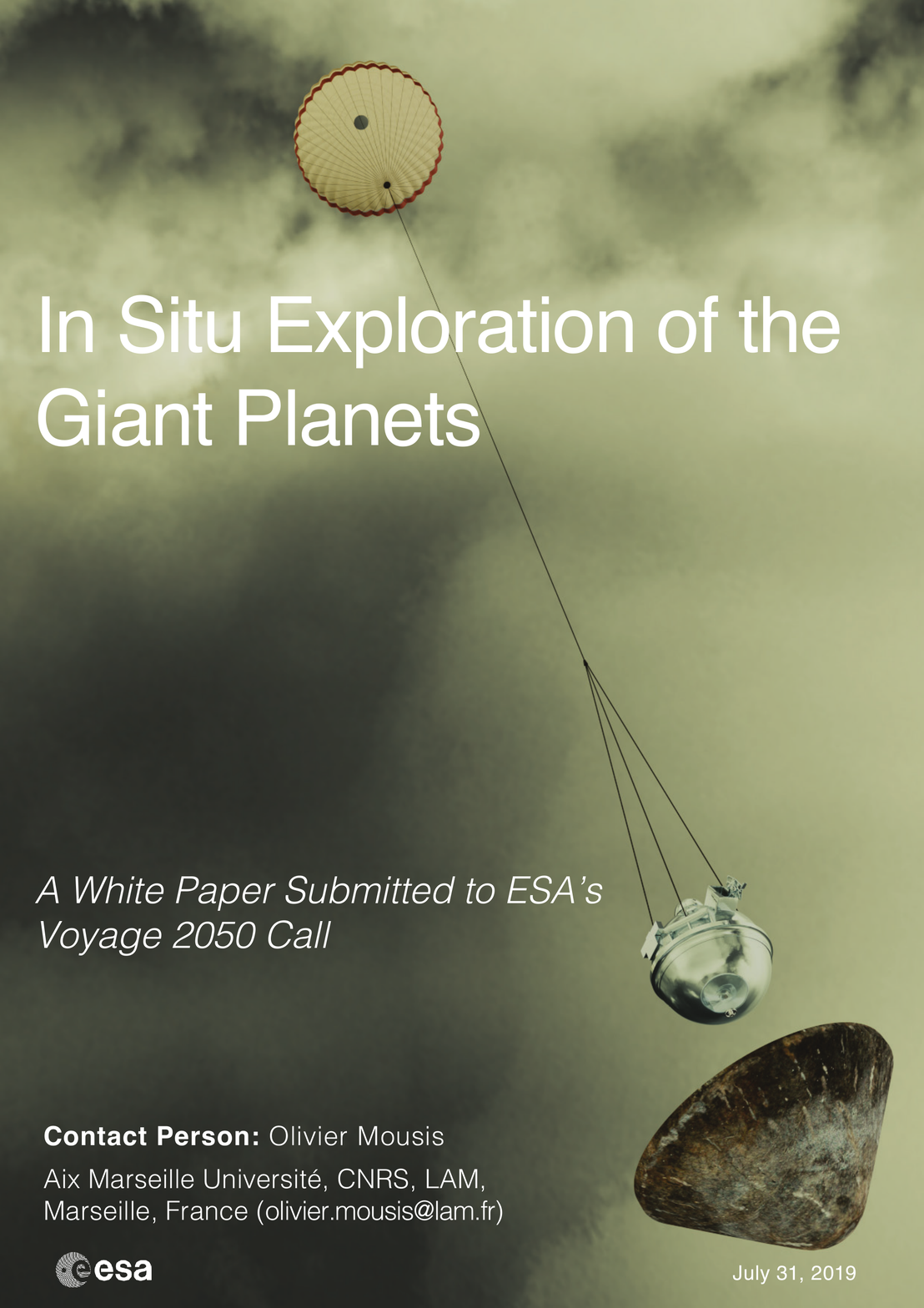}

\flushbottom 

\maketitle 

\tableofcontents 

\thispagestyle{empty} 


\section{Context} 
\subsection{Why In Situ Measurements in Giant Planets?}

Giant planets contain most of the mass and the angular momentum of our planetary system and must have played a significant role in shaping its large scale architecture and evolution, including that of the smaller, inner worlds \cite{Go05}. Furthermore, the formation of the giant planets affected the timing and efficiency of volatile delivery to the Earth and other terrestrial planets \cite{Ch01}. Therefore, understanding giant planet formation is essential for understanding the origin and evolution of the Earth and other potentially habitable environments throughout our solar system. The origin of the giant planets, their influence on planetary system architectures, and the plethora of physical and chemical processes at work within their atmospheres make them crucial destinations for future exploration. Since Jupiter and Saturn have massive envelopes essentially composed of hydrogen and helium and (possibly) a relatively small core, they are called gas giants. Uranus and Neptune also contain hydrogen and helium atmospheres but, unlike Jupiter and Saturn, their H$_2$ and He mass fractions are smaller (5--20\%). They are called ice giants because their density is consistent with the presence of a significant fraction of ices/rocks in their interiors. Despite this apparent grouping into two classes of giant planets, the four giant planets likely exist on a continuum, each a product of the particular characteristics of their formation environment. {\it Comparative planetology of the four giants in the solar system} is therefore essential to reveal the potential formational, migrational, and evolutionary processes at work during the early evolution of the early solar nebula. As discussed below, {\it in situ} exploration of the four giants is the means to address this theme.

\begin{figure*}[ht]
\centering 
\includegraphics[width=15cm]{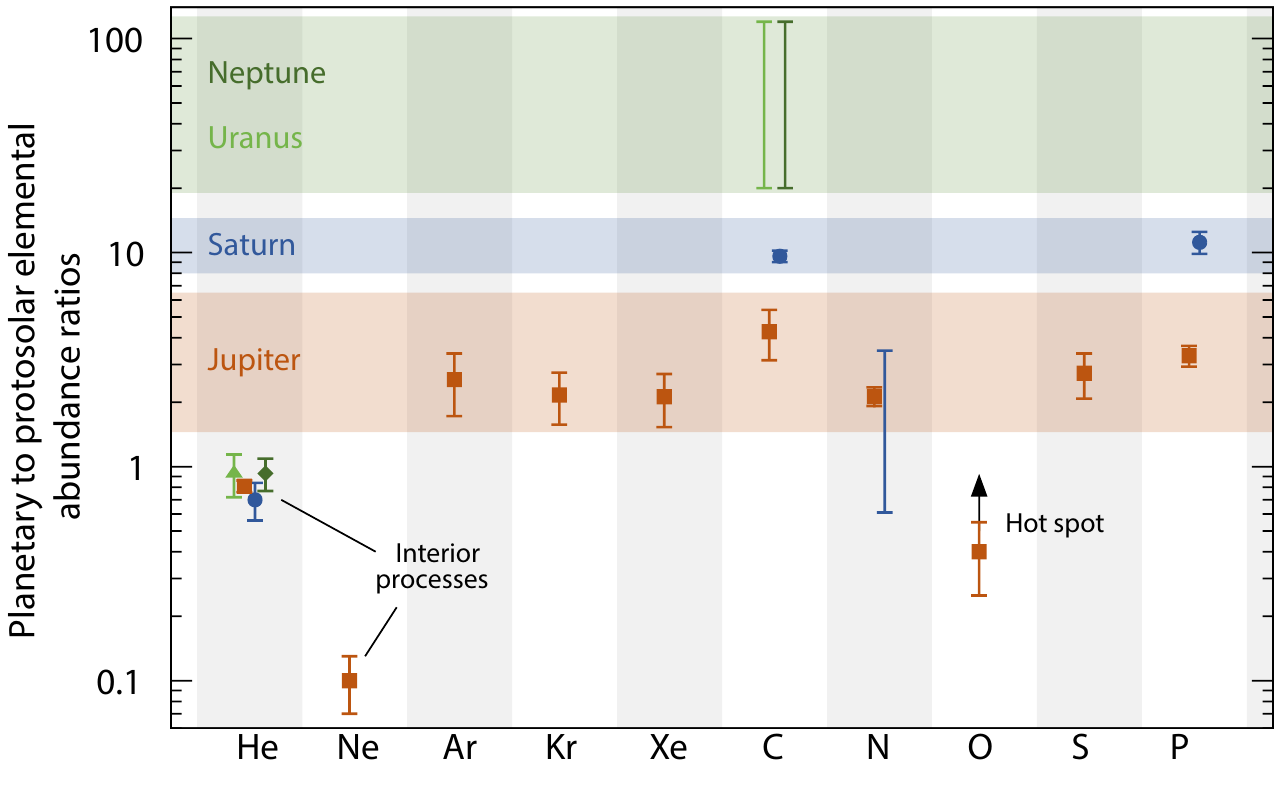}
\caption{Enrichment factors (with respect to the protosolar value) of noble gases and heavy elements measured in Jupiter, Saturn, Uranus, and Neptune. Error bars, central values and planets share the same color codes (see \cite{Mo18} for references).}
\label{fig:enri}
\end{figure*}

Much of our understanding of the origin and evolution of the outer planets comes from remote sensing by necessity. However, {\it the efficiency of this technique has limitations when used to study the bulk atmospheric composition that is crucial to the understanding of planetary origin}, primarily due to degeneracies between the effects of temperatures, clouds and abundances on the emergent spectra, but also due to the limited vertical resolution. In addition, many of the most abundant elements are locked away in a condensed phase in the upper troposphere, hiding the main volatile reservoir from the reaches of remote sensing. It is only by penetrating below the ``visible'' weather layer that we can sample the deeper troposphere where those elements are well mixed. A remarkable example of the superiority of {\it in situ} probe measurements is illustrated by the exploration of Jupiter, where key measurements such as the determination of the abundances of noble gases and the precise measurement of the helium mixing ratio have only been possible through {\it in situ} measurements by the Galileo probe \cite{Ow99}.

The Galileo probe measurements provided new insights into the formation of the solar system. For instance, they revealed the
unexpected enrichments of Ar, Kr and Xe with respect to their solar abundances (see Figure \ref{fig:enri}), which suggested that the planet accreted icy planetesimals formed at temperatures possibly below $\sim$50~K to enable the trapping of these noble gases. Another remarkable result was the determination of the Jovian helium abundance using a dedicated instrument aboard the Galileo probe \cite{vo98} with an accuracy of 2\%. Such an accuracy on the He/H$_2$ ratio is impossible to derive from remote sensing, irrespective of the giant planet being considered, and yet precise knowledge of this ratio is crucial for the understanding of giant planet interiors and thermal evolution. The Voyager mission has already shown that these ratios are far from being identical in the gas and icy giants, which presumably result from different thermal histories and internal processes at work. Another important result obtained by the mass spectrometer onboard the Galileo probe was the determination of the $^{14}$N/$^{15}$N ratio, which suggested that nitrogen present in Jupiter today originated from the solar nebula essentially in the form of N$_2$ \cite{Ow01}. The Galileo science payload unfortunately could not probe to pressure levels deeper than 22 bar, precluding the determination of the H$_2$O abundance at levels representative of the bulk oxygen enrichment of the planet. Furthermore, the probe descended into a region depleted in volatiles and gases by unusual ``hot spot'' meteorology \cite{Or98,Wo04}, and therefore its measurements are unlikely to represent the bulk planetary composition. Nevertheless, the Galileo probe measurements were a giant step forward in our understanding of Jupiter. However, with only a single example of a giant planet measurement, one must wonder to what extent from the measured pattern of elemental and isotopic enrichments, the chemical inventory and formation processes at work in our solar system are truly understood. {\it In situ} exploration of giant planets is the only way to firmly characterize their composition. In this context, one or several entry probes sent to the atmosphere of any of the other giant planets of our solar system is the next natural step beyond Galileo's {\it in situ} exploration of Jupiter, the remote investigation of its interior and gravity field by the Juno mission, and the Cassini spacecraft's orbital reconnaissance of Saturn.

{\it In situ} exploration of Saturn, Uranus or Neptune's atmospheres addresses two broad themes. First, the formation history of our solar system and second, the processes at play in planetary atmospheres. Both of these themes are discussed throughout this White Paper. Both themes have relevance far beyond the leap in understanding gained about an individual giant planet: the stochastic and positional variances produced within the solar nebula, the depth of the zonal winds, the propagation of atmospheric waves, the formation of clouds and hazes and disequilibrium processes of photochemistry and vertical mixing are common to all planetary atmospheres, from terrestrial planets to gas and ice giants and from brown dwarfs to hot exoplanets.

\subsection{Entry Probes in the Voyage 2050 Programme}

The {\it in situ} exploration of Saturn, Uranus, and/or Neptune fits perfectly within the ambitious scope of the ESA Voyage 2050 Programme. A Saturn entry probe proposal has already been submitted to the ESA M4 and M5 calls in 2015 and 2016, respectively. Experience from these submissions shows that the development of entry probes match well the envelope allocated to ESA M-class missions provided that the carrier is provided by another space agency. Selection for phase A failed during the M4 and M5 evaluations because of the lack of availability of a NASA carrier at the envisaged launch epoch. An ideal combination would be a partnership between ESA and NASA in which ESA provides an entry probe as an important element of a more encompassing NASA New Frontiers or Flagship mission toward Saturn, Uranus, or Neptune. A joint NASA-ESA Ice Giant Study Science Definition Team (SDT) has been set in 2016-2017 to investigate the best mission scenarios dedicated to the exploration of Uranus and Neptune in terms of science return \cite{Ho19}. The conclusions of the study outline the high priority of sending an orbiter and atmospheric probe to at least one of the ice giants. The mission architectures assessed by the 2017 NASA SDT showed that 2030--34 were the optimal launch windows for Uranus, but it would be even earlier (2029--30) for Neptune, depending on the use of Jupiter for a gravity assist. An internal ESA study led at the end of 2018 (ESA M* Ice Giant CDF study 1) shows that the technology is available in Europe to provide a probe to NASA\footnote{http://sci.esa.int/future-missions-department/61307-cdf-study-report-ice-giants/} in the framework of a joint mission. Apart the DragonFly mission dedicated to the exploration of Titan and recently selected by NASA for launch in 2026, future New Frontiers proposals could also be devoted to the {\it in situ} exploration of Saturn \cite{Ba18z}. The selection of such proposals could create an ideal context for ESA to contribute an entry probe to NASA. Under those circumstances, the dropping of one or several probes could be envisaged in the atmosphere of Saturn.


\section{Science Themes}
\label{science}
\subsection{Elemental and Isotopic Composition as a Window on the Giant Planets Formation}

The giant planets in the solar system formed 4.55 Gyr ago from the same material that engendered the Sun and the entire solar system.  
Protoplanetary disks, composed of gas and dust, are almost ubiquitous when stars form, but their typical lifetimes do not exceed a few million years. This implies that the gas giants Jupiter and Saturn had to form rapidly to capture their hydrogen and helium envelopes, more rapidly than the tens of millions of years needed for terrestrial planets  to reach their present masses \cite{Po96,Al05a,Al05b}. Due to formation at fairly large radial distances from the Sun, where the solid surface density is low, the ice giants Uranus and Neptune had longer formation timescales (slow growth rates) and did not manage to capture large amounts of hydrogen and helium before the disk gas dissipated \cite{Do10,He14a}. As a result, the masses of their gaseous envelopes are small compared to their ice/rock cores. A comparative study of the properties of these giant planets thus gives information on spatial gradients in the physical and chemical properties of the solar nebula as well as on stochastic effects that led to the formation of the solar system. Data on the composition and structure of the giant planets, which hold more than 95\% of the mass of the solar system outside of the Sun, remain scarce, despite the importance of such knowledge. The formation of giant planets is now largely thought to have taken place via the {\it core accretion model} in which a dense core is first formed by accretion and the hydrogen-helium envelope is captured after a critical mass is reached \cite{Mi78,Po96}. When the possibility of planet migration is included \cite{Li86,wa97}, such a model may be able to explain the orbital properties of exoplanets, although lots of unresolved issues remain \cite{Id04,Mor12}. An alternative giant planets formation scenario is also the {\it gravitational instability model} \cite{Bo97,Bo01}, in which the giant planets form from the direct contraction of a gas clump resulting from local gravitational instability in the disk. 

In the following, we briefly review the interior models, as well as the chemical and isotopic compositions of the four giants of our solar system. We also investigate the enrichment patterns that could be derived from {\it in situ} measurements by entry probes in the giant planets atmospheres to derive hints on their formation conditions. We finally summarize the key observables accessible to an atmospheric probe to address the scientific issues to the formation and evolution of the giant planets.

\paragraph{Interior Models} Interior models for the present state of the planets serve as a link between the formation scenarios outlined above and observations. Notably, recent interior models of Jupiter that fit the gravity data observed by NASA's current Juno spacecraft are consistent with a deep interior that is highly enriched in heavy elements up to about 60\% of the planet's radius. Comparison of such interior models to models of Jupiter's formation and evolution implies that the deep interior still retains a memory of the infall of planetesimals  at the time of formation \cite{Va18}. In that scenario, accretion of heavy elements into the growing envelope led to persistent compositional gradients that are still inhibiting efficient convection and mixing. However, Jupiter interior models greatly differ in the predicted amount of heavy elements in the atmosphere, which is accessible to observations. Predictions range from less than 1 $\times$ solar \cite{Wa17} over 1--2 $\times$ solar \cite{Ne17} to  $\sim$6 $\times$ solar \cite{Va18}. These differences are mostly due to uncertainties in the H/He Equation of State (EOS) and can be compared with the atmospheric abundances of elements measured in giant planets atmospheres provided they are representative of the bulk envelope. Such comparisons are highly valuable for constraining formation models and for a better understanding of the interplay between the H/He EOS and the structure of gaseous planets. In the case of Jupiter, at minimum the heavy noble gas abundances measured by the Galileo probe serve that purpose. NASA's Juno mission currently tries to obtain the H$_2$O abundance. However, the microwave spectra are highly influenced by the NH$_3$ abundances rendering the quantitative assessment through remote sensing difficult. Bulk heavy element masses in Jupiter are estimated to range from $\sim$25 \Mearth \cite{Wa17} to over $\sim$32 \Mearth \cite{Ne17} up to 40 \Mearth \cite{Va18}. 

In the case of Saturn, the mass of heavy elements can vary between 0 and $\sim$7 \Mearth~in the envelope, and between 5 and 20~\Mearth~in the core~\cite{He13}. Similar to Jupiter, potential compositional inhomogeneities in Saturn could be the outcome of the formation process \cite{Po96} and/or the erosion of a primordial core that could mix with the surrounding metallic hydrogen \cite{Wi12a,Wi12b}. In addition, it is possible that double diffusive convection occurs in the interior of Saturn \cite{Le12,Le13}. If a molecular weight gradient is maintained throughout the planetary envelope, double-diffusive convection would take place, and the thermal structure would be very different from the one that is generally assumed using adiabatic models, with much higher center temperatures and a larger fraction of heavy elements. In this case, the planetary composition can vary substantially with depth and therefore, a measured composition of the envelope would not represent the overall composition. While standard interior models of Saturn assumed three layers and similar constraints in terms of the helium to hydrogen ratio, they can differ in the assumption on the distribution of heavy elements within the planetary envelope: homogeneous distribution of heavy elements apart from helium, which is depleted in the outer envelope due to helium rain \cite{Sa04,He13} or interior structure models allowing the abundance of heavy elements to be discontinuous between the molecular and the metallic envelope \cite{Fo10,Ne13}. At present, it is not clear whether there should be a discontinuity in the composition of heavy elements, and this question remains open.

Because of the scarcity of data, the interiors of Uranus and Neptune are even less constrained.  Improved gravity field data derived from long-term observations of the planets' satellite motions suggests however that Uranus and Neptune could have different distributions of heavy elements \cite{Ne13}. These authors estimate that the bulk masses of heavy elements are $\sim$12.5 \Mearth~for Uranus and $\sim$14 \Mearth~for Neptune. They also find that Uranus would have an outer envelope with a few times the solar metallicity which transitions to a heavily enriched ($\sim$90\% of the mass in heavy elements) inner envelope at 0.9 planet's radius. In the case of Neptune, this transition is found to occur deeper inside at 0.6 planet's radius and accompanied with a more moderate increase in metallicity. Direct access to heavy materials within giant planet cores to constrain these models is impossible, so we must use the composition of the well-mixed troposphere to infer the properties of the deep interiors. It is difficult for remote sounding to provide the necessary information because of a lack of sensitivity to the atmospheric compositions beneath the cloudy, turbulent and chaotic weather layer. These questions must be addressed by {\it in situ} exploration, even if the NASA Juno mission is successful in addressing some of them remotely at Jupiter.

\paragraph{Giant Planets Composition} The abundances of most significant volatiles measured at Jupiter, Saturn, Uranus, and Neptune are summarized in Tables \ref{table1} and \ref{table2}. The composition of giant planets is diagnostic of their formation and evolution history. Measuring their heavy element, noble gas, and isotope abundances reveals the physico-chemical conditions and processes that led to formation of the planetesimals that eventually fed the forming planets \cite{Ow99,Ga01,He01}. Heavy element abundances can be derived through a variety of remote sensing techniques such as spectroscopy. However, the most significant step forward regarding our knowledge of giant planet internal composition was achieved with the {\it in situ} descent of the Galileo probe into the atmosphere of Jupiter \cite{yo98,Fo98,Ra98,At98,Sr98,Ni98,vo98}. The various experiments enabled the determination of the He/H$_2$ ratio with a relative accuracy of 2\% \cite{vo98}, of several heavy element abundances and of noble gases abundances \cite{Ni98,At99,Wo04}. These measurements have paved the way to a better understanding of Jupiter's formation and evolution. For example, neon in Jupiter's atmospheres has been found to be the most strongly depleted element. Its depletion, in contrast to the measured enrichments in Ar, Kr, Xe, is attributed to the helium rain in Jupiter \cite{Wi10}. It would be very valuable to have measurements of the heavy noble gases in any other giant planet. For Saturn, we would expect a similarly strong depletion in neon as in Jupiter as a result of deep atmospheric helium rain whereas in Uranus and Neptune depletion in He and Ne is not expected. This is because their deep interiors are mostly made of ices, implying that He is rare there and does not rain out. {\it In situ} measurements in all of these planets atmospheres would thus allow us to test these assumptions and to offer a diagnostic tool of the behavior of H/He at high pressures in giant planets. The uniform enrichment observed in the Galileo probe data (see Figure \ref{fig:enri}) tends to favor a {\it core accretion} scenario for Jupiter (e.g. \cite{Al05a,Gu05}), even if the gravitational capture of planetesimals by the proto-Jupiter formed via {\it gravitational instability} may also explain the observed enrichments \cite{He06}. On the other hand, the condensation processes that formed the protoplanetary ices remain uncertain, because the Galileo probe failed to measure the deep abundance of oxygen by diving into a dry area of Jupiter \cite{At03}. Achieving this measurement by means of remote radio observations is one of the key and most challenging goals of the Juno mission \cite{Ma07,He14b}, currently in orbit around Jupiter. At Saturn, the data on composition are scarcer (see Figure \ref{fig:enri}) and have mostly resulted from Voyager~2 measurements and intense observation campaigns with the Cassini orbiter. The He abundance is highly uncertain \cite{Co84,Co00,Ac16}, and only the abundances of N, C, and P, have been quantified \cite{Cou84,Da96,Fl07,Fl09a,Fl09b}. This scarcity of essential data is the main motivation for sending an atmospheric probe to Saturn and was the core of several mission proposals submitted to ESA and NASA calls over the last decade \cite{Mo14a,Mo16,At16}. Uranus and Neptune are the most distant planets in our Solar System. Their apparent size in the sky is roughly a factor of 10 smaller than Jupiter and Saturn, which makes telescopic observations from Earth much more challenging in terms of detectability. This distance factor is probably also the reason why space agencies have not yet sent any new flyby or orbiter mission to either of these planets since Voyager 2. As a consequence, the knowledge of their bulk composition is dramatically poor (see Figure \ref{fig:enri}), resulting in a very limited understanding of their formation and evolution. Improving this situation needs ground-truth measurements that can only be carried out in these distant planets by an atmospheric probe, similarly to the Galileo probe at Jupiter.

\begin{table*}[h]
\begin{center}
\caption[]{Elemental abundances in Jupiter, Saturn, Uranus and Neptune, as derived from upper tropospheric composition \label{table1}}
\small\begin{tabular}{lllll}
\toprule
Elements	& Jupiter 								& Saturn						& Uranus						& Neptune 										\\
\midrule
He/H	$^{(1)}$ 		& ($7.85 \pm 0.16) \times 10^{-2}$		& $(6.75 \pm 1.25) \times 10^{-2}$		&  $(8.88 \pm 2.00) \times 10^{-2}$	& $(8.96 \pm 1.46) \times 10^{-2}$				\\
Ne/H$^{(2)}$ 		& ($1.240 \pm 0.014) \times 10^{-5}$		& --								&  --							& --										\\
Ar/H$^{(3)}$ 		& ($9.10 \pm 1.80) \times 10^{-6}$		& --								& --							& --										\\
Kr/H$^{(4)}$ 		& ($4.65 \pm 0.85) \times 10^{-9}$		& --								& --							& --										\\
Xe/H	$^{(5)}$	 	& ($4.45 \pm 0.85) \times 10^{-10}$		& --								& --							& --										\\
C/H$^{(6)}$ 		& ($1.19 \pm 0.29) \times 10^{-3}$		& $(2.65 \pm 0.10) \times 10^{-3}$		& $(0.6 - 3.2) \times 10^{-2}$ 		& $(0.6 - 3.2) \times 10^{-2}$					\\
N/H$^{(7)}$ 		& ($3.32 \pm 1.27) \times 10^{-4}$		& $(0.50 - 2.85) \times 10^{-4}$			& --							& --										\\
O/H$^{(8)}$ 		& ($2.45 \pm 0.80) \times 10^{-4}$		& --    							& --							& --							  			 \\
S/H$^{(9)}$ 		& ($4.45 \pm 1.05) \times 10^{-5}$		& --                             				& ($5 - 12.5) \times 10^{-6}$		& ($2.0 - 6.5) \times 10^{-6}$ 					\\
P/H$^{(10)}$ 		& ($1.08 \pm 0.06) \times 10^{-6}$		& ($3.64 \pm 0.24) \times 10^{-6}$		& --							& --										\\			
\bottomrule
\end{tabular}\\
$^{(1)}$ \cite{vo98,Ni98} for Jupiter, \cite{Co00,At16} for Saturn, \cite{Co87} for Uranus and \cite{Bu03} for Neptune. We only consider the higher value of the uncertainty on He in the case of Neptune. $^{(2-5)}$ \cite{Ma00} for Jupiter. $^{(6)}$ \cite{Wo04} for Jupiter, \cite{Fl09a} for Saturn, \cite{Li87,ba95,Ka09,Sr14} for Uranus, \cite{Li90,ba95,ka11} for Neptune. $^{(7)}$ \cite{Wo04} for Jupiter, \cite{Fl11} for Saturn (N/H range derived from the observed range of 90--500 ppm of NH$_3$). $^{(8)}$ \cite{Wo04} for Jupiter (probably a lower limit, not representative of the bulk O/H). $^{(9)}$ \cite{Wo04} for Jupiter, lower limits for Uranus \cite{Ir18} and Neptune \cite{Ir19}. $^{(10)}$ \cite{Fl09b} for Jupiter and Saturn.
\label{table1}
\end{center}
\end{table*}

\begin{table*}[h]
\begin{center}
\caption[]{Ratios to protosolar values in the upper tropospheres of Jupiter, Saturn, Uranus and Neptune \label{table2}}
\small\begin{tabular}{lllll}
\toprule
Elements	& Jupiter/Protosolar			& Saturn/Protosolar			&  Uranus/Protosolar			&  Neptune/Protosolar\\
\midrule
He/H		& $0.81 \pm 0.05$			& $0.70 \pm 0.14$			& $0.93 \pm 0.21$			& $0.93 \pm 0.16$			\\
Ne/H		& $0.10 \pm 0.03$			& --						& --						& --						\\
Ar/H		& $2.55 \pm 0.83$			& --						& --						& --						\\
Kr/H		& $2.16 \pm 0.59$			& --						& --						& --						\\
Xe/H		& $2.12 \pm 0.59$			& --						& --						& --						\\
C/H		& $4.27 \pm 1.13$			& $9.61 \pm 0.59$			& $\sim$20 -- 120		        & $\sim$20 -- 120   	                 \\
N/H		& $4.06 \pm 2.02$			& 0.61 -- 3.48				& --						& --						\\
O/H		& $0.40 \pm 0.15$ (hotspot)	& --    					& --						& --						\\
S/H		& $2.73 \pm 0.65$			& --    					& 0.32 - 0.80				& 0.13 - 0.42				\\
P/H		& $3.30 \pm 0.37$			& $11.17 \pm 1.31$			& --						& --						\\			
\bottomrule
\end{tabular}\\
Error is defined as ($\Delta$E/E)$^2$ =  ($\Delta$X/X$_{\rm planet}$)$^2$ + ($\Delta$X/X$_{\rm Protosun}$)$^2$. The ratios only refer to the levels where abundance measurements have been performed, i.e. in the upper tropospheres and are not automatically representative of deep interior enrichments. This is especially true if the deep interior contain a significant fraction of another element (e.g. oxygen in Uranus and Neptune, according to models). Moreover, the helium value was computed for pure H$_2$/He mixtures (i.e. the upper tropospheric CH$_4$ has not been accounted for), because CH$_4$ is condensed at 1\,bar where He is measured. Protosolar abundances are taken from \cite{Lo09}.
\label{table2}
\end{center}
\end{table*}

\paragraph{Isotopic Measurements}  Table \ref{table3} represents the isotopic ratio measurements realized in the atmospheres of the four giant planets of our solar system. The case of D/H is interesting and would deserve further measurements with smaller errors. Because deuterium is destroyed in stellar interiors and transformed into $^3$He, the D/H value presently measured in Jupiter's atmosphere is estimated to be larger by some 5--10\% than the protosolar value. This slight enrichment would have resulted from a mixing of nebular gas with deuterium-rich ices during the planet's formation. For Saturn, the contribution of deuterium-rich ices in the present D/H ratio could be higher (25--40\%). The deuterium enrichment as measured by \cite{Fe13} in Uranus and Neptune has been found to be very similar between the two planets, and its supersolar value also suggests that significant mixing occurred between the protosolar H$_2$ and the H$_2$O ice accreted by the planets. Assuming that the D/H ratio in H$_2$O ice accreted by Uranus and Neptune is cometary (1.5--3 $\times $10$^{-4}$), \cite{Fe13} found that 68--86\% of the heavy component consists of rock and 14--32\% is made of ice, values suggesting that both planets are more rocky than icy, assuming that the planets have been fully mixed. Alternatively, based on these observations, \cite{Al14} suggested that, if Uranus and Neptune formed at the carbon monoxide line in the protosolar nebula (PSN), then the heavy elements accreted by the two planets would mostly consists of a mixture of CO and H$_2$O ices, with CO being by far the dominant species. This scenario assumes that the accreted H$_2$O ice presents a cometary D/H and allows the two planets to remain ice-rich and O-rich while providing D/H ratios consistent with the observations. Deeper sounding of Saturn, Uranus, and Neptune's atmospheres with an atmospheric probe, should allow investigating the possibility of isotopic fractionation with depth. The measurement of the D/H ratio in Saturn, Uranus and Neptune should be complemented by a precise determination of $^3$He/$^4$He in their atmospheres to provide further constraints on the protosolar D/H ratio, which remains relatively uncertain. The protosolar D/H ratio is derived from $^3$He/$^4$He measurements in the solar wind corrected for changes that occurred in the solar corona and chromosphere subsequent to the Sun's evolution, and to which the primordial $^3$He/$^4$He is subtracted \cite{Ge98}. This latter value is currently derived from the ratio observed in meteorites or in Jupiter's atmosphere. The measurement of $^3$He/$^4$He in Uranus and/or Neptune atmospheres would therefore complement the Jupiter value and the scientific impact of the protosolar D/H derivation.

The $^{14}$N/$^{15}$N ratio presents large variations in the different planetary bodies in which it has been measured and, consequently, remains difficult to interpret. The analysis of Genesis solar wind samples \cite{Ma11} suggests a $^{14}$N/$^{15}$N ratio of 441 $\pm$ 5, which agrees with the remote sensing \cite{Fo00} and {\it in situ} \cite{Wo04} measurements made in Jupiter's atmospheric ammonia, and the lower limit derived from ground-based mid-infrared observations of Saturn's ammonia absorption features \cite{Fl14}. The two $^{14}$N/$^{15}$N measurements made in Jupiter and Saturn suggest that primordial N$_2$ was probably the main reservoir of the present NH$_3$ in their atmospheres \cite{Ow01,Mo14a,Mo14b}. On the other hand, Uranus and Neptune are mostly made of solids (rocks and ices) \cite{Gu05} that may share the same composition as comets. N$_2$/CO has been found strongly depleted in comet 67P/Churyumov-Gerasimenko \cite{Ru15}, i.e. by a factor of $\sim$25.4 compared to the value derived from protosolar N and C abundances. This confirms the fact that N$_2$ is a minor nitrogen reservoir compared to NH$_3$ and HCN in this body \cite{Le15}, and probably also in other comets \cite{Bo04}. In addition, $^{14}$N/$^{15}$N has been measured to be 127 $\pm$ 32 and 148 $\pm$ 6 in cometary NH$_3$ and HCN respectively \cite{Ro14,Ma09}.   Assuming that Uranus and Neptune have been accreted from the same building blocks as those of comets, then one may expect a $^{14}$N/$^{15}$N ratio in these two planets close to cometary values, and thus quite different from the Jupiter and Saturn values. Measuring $^{14}$N/$^{15}$N in the atmospheres of Uranus and Neptune would provide insights about the origin of the primordial nitrogen reservoir in these planets. Moreover, measuring this ratio in different species would enable us to constrain the relative importance of the chemistry induced by galactic cosmic rays and magnetospheric electrons (see \cite{Do18} for an example in Titan).

The isotopic measurements of carbon, oxygen and noble gas (Ne, Ar, Kr, and Xe) isotopic ratios should be representative of their primordial values. For instance, only little variations are observed for the $^{12}$C/$^{13}$C ratio in the solar system irrespective of the body and molecule in which it has been measured. Table \ref{table3} shows that both ratios measured in the atmospheres of Jupiter and Saturn are consistent with the terrestrial value of 89. A new {\it in situ} measurement of this ratio in Uranus and/or Neptune should be useful to confirm whether their carbon isotopic ratio is also telluric. 

The oxygen isotopic ratios also constitute interesting measurements to be made in Uranus' and Neptune's atmospheres. The terrestrial $^{16}$O/$^{18}$O and $^{16}$O/$^{17}$O isotopic ratios are 499 and 2632, respectively \cite{As09}. At the high accuracy levels achievable with meteoritic analysis, these ratios present some small variations (expressed in $\delta$ units, which are deviations in part per thousand). Measurements performed in comets \cite{Bo12}, far less accurate, match the terrestrial $^{16}$O/$^{18}$O value. The $^{16}$O/$^{18}$O ratio has been found to be $\sim$380 in Titan's atmosphere from Herschel SPIRE observations but this value may be due to some fractionation process \cite{Co11,Lo17}. On the other hand, \cite{Se16} found values consistent with the terrestrial ratios in CO with ALMA. The only $^{16}$O/$^{18}$O measurement made so far in a giant planet was obtained from ground-based infrared observations in Jupiter's atmosphere and had a too large uncertainty to be interpreted in terms of 1--3 times the terrestrial value \cite{No95}.

\begin{table*}
\begin{center}
\caption[]{Isotopic ratios measured in Jupiter, Saturn, Uranus and Neptune}
\small\begin{tabular}{lcccccc}
\toprule
Isotopic ratio							& Jupiter							& Saturn								&Uranus							& Neptune						\\
\midrule
D/H (in H$_2$)$^{(1)}$					& (2.60 $\pm$ 0.7) $\times$ 10$^{-5}$	& $1.70 ^{+0.75}_{-0.45}$ $\times$ 10$^{-5}$	& (4.4 $\pm$ 0.4) $\times$ 10$^{-5}$		& (4.1 $\pm$ 0.4) $\times$ 10$^{-5}$		\\
$^3$He/$^4$He$^{(2)}$					& (1.66 $\pm$ 0.05) $\times$ 10$^{-4}$	& --									& --								& --  								\\
$^{12}$C/$^{13}$C (in CH$_4$)$^{(3)}$		& 92.6$^{+4.5}_{-4.1}$				& 91.8$^{+8.4}_{-7.8}$					& -- 								& -- 								\\
$^{14}$N/$^{15}$N (in NH$_3$)$^{(4)}$		& 434.8$^{+65}_{-50}$				& $>357$ 								& --								& --								\\
$^{20}$Ne/$^{22}$Ne$^{(5)}$				& 13 $\pm$ 2						& --									& --								& -- 								\\
$^{36}$Ar/$^{38}$Ar$^{(6)}$				& 5.6	 $\pm$ 0.25					& --									& --								& -- 								\\
$^{136}$Xe/total Xe$^{(7)}$				& 0.076 $\pm$	0.009				& --									& --								& --  								\\
$^{134}$Xe/total Xe$^{(8)}$				& 0.091 $\pm$ 0.007					& --									& --								& --  								\\
$^{132}$Xe/total Xe$^{(9)}$				& 0.290 $\pm$ 0.020					& --									& --								& -- 								\\
$^{131}$Xe/total Xe$^{(10)}$				& 0.203 $\pm$ 0.018					& --									& --								& --  								\\
$^{130}$Xe/total Xe$^{(11)}$				& 0.038 $\pm$ 0.005					& --									& --								& -- 								\\
$^{129}$Xe/total Xe$^{(12)}$				& 0.285 $\pm$ 0.021					& --									& --								& --  								\\
$^{128}$Xe/total Xe$^{(13)}$				& 0.018 $\pm$ 0.002					& --									& --								& -- 								\\
\bottomrule
\end{tabular}\\
$^{(1)}$ \cite{Ma98} for Jupiter, \cite{Le01} for Saturn, \cite{Fe13} for Uranus and Neptune. $^{(2)}$ \cite{Ma98} for Jupiter. $^{(3)}$ \cite{Ni98} for Jupiter, \cite{Fl09a} for Saturn. $^{(4)}$ \cite{Wo04} for Jupiter, \cite{Fl14} for Saturn. $^{(5-13)}$ \cite{Ma00} for Jupiter.
\label{table3}
\end{center}
\end{table*}

\begin{figure*}[ht]
\begin{center}
\includegraphics[width=16cm]{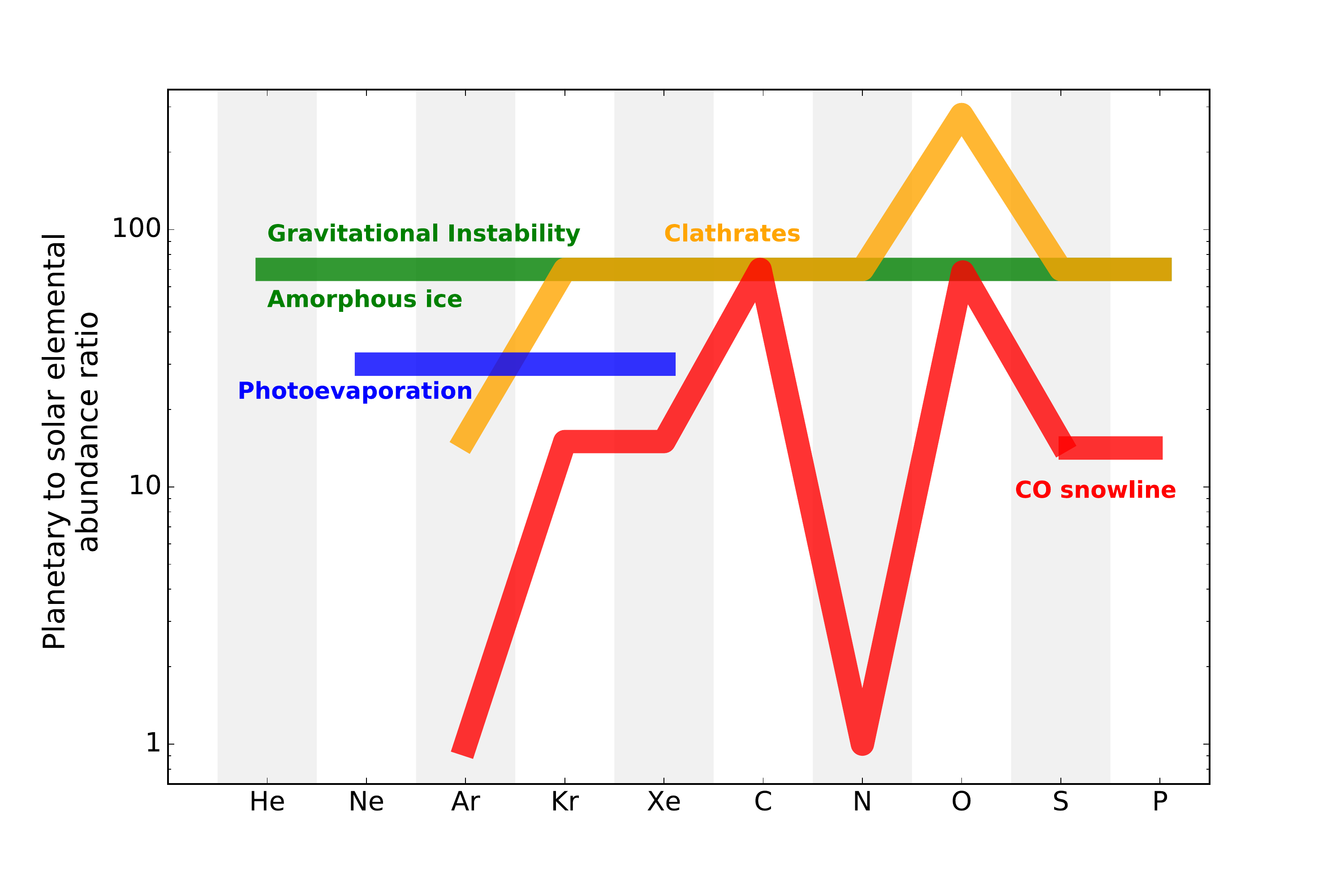}
\caption{Qualitative differences between the enrichments in volatiles predicted in Uranus and Neptune predicted by the different formation scenarios (calibrations based on the carbon determination). The resulting enrichments for the different volatiles are shown in green (gravitational instability model and amorphous ice), orange (clathrates), blue (photoevaporation) and red (CO snowline).}
\label{enri_pred}
\end{center}
\end{figure*} 

\paragraph{Formation Models and Enrichment Patterns in Giant Planets} Direct or indirect measurements of the volatile abundances in the atmospheres of Saturn, Uranus and Neptune are key for deciphering their formation conditions in the PSN. In what follows, we present the various models and their predictions regarding enrichments in the giants. Figure \ref{enri_pred} summarizes the predictions of the various models in the cases of Uranus and Neptune.

\begin{itemize}
\item {\it Gravitational Instability Model.} This formation scenario is associated with the photoevaporation of the giant planets envelopes by a nearby OB star and settling of dust grains prior to mass loss \cite{Bo02}. It implies that O, C, N, S, Ar, Kr and Xe elements should all be enriched by a similar factor relative to their protosolar abundances in the envelopes, assuming mixing is efficient. Despite the fact that interior models predict that a metallicity gradient may increase the volatile enrichments at growing depth in the planet envelopes \cite{Ne13}, there is no identified process that may affect their relative abundances in the ice giant envelopes, if the sampling is made at depths below the condensation layers of the concerned volatiles and if thermochemical equilibrium effects are properly taken into account. The assumption of homogeneous enrichments for O, C, N, S, Ar, Kr and Xe, relative to their protosolar abundances, then remains the natural outcome of the formation scenario proposed by \cite{Bo02}.

\item {\it Core Accretion and Amorphous Ice.} In the case of the {\it core accretion} model, because the trapping efficiencies of C, N, S, Ar, Kr and Xe volatiles are similar at low temperature in amorphous ice \cite{Ow99,Bar07}, the delivery of such solids to the growing giant planets is also consistent with the prediction of homogeneous enrichments in volatiles relative to their protosolar abundances in the envelopes, still under the assumption that there is no process leading to some relative fractionation between the different volatiles. 

\item{\it Core Accretion and Clathrates} In the {\it core accretion} model, if the volatiles were incorporated in clathrate structures in the PSN, then their propensities for such trapping would strongly vary from a species to another. For instance, Xe, CH$_4$ and CO$_2$ are easier clathrate formers than Ar or N$_2$ because their trapping temperatures are higher at PSN conditions, assuming protosolar abundances for all elements \cite{Mo10}. This competition for trapping is crucial when the budget of available crystalline water is limited and does prevent the full clathration of the volatiles present in the PSN \cite{Ga01,Mo12,Mo14b}. However, if the O abundance is 2.6 times protosolar or higher at the formation locations of Uranus and Neptune's building blocks and their formation temperature does not exceed $\sim$45K, then the abundance of crystalline water should be high enough to fully trap all the main C, N, S and P--bearing molecules, as well as Ar, Kr and Xe \cite{Mo14b}. In this case, all elements should present enrichments comparable to the C measurement, except for O and Ar, based on calculations of planetesimals compositions performed under those conditions \cite{Mo14b}. The O enrichment should be at least $\sim$4 times higher than the one measured for C in the envelopes of the ice giants due to its overabundance in the PSN. In contrast, the Ar enrichment is decreased by a factor of $\sim$4.5 compared to C, due to its very poor trapping at 45 K in the PSN (see Figure \ref{enri_pred}). We refer the reader to \cite{Mo14b} for further details about the calculations of these relative abundances. 

\item {\it Photoevaporation Model.} An alternative scenario is built upon the ideas that (i) Ar, Kr and Xe were homogeneously adsorbed at very low temperatures ($\sim$20--30 K) at the surface of amorphous icy grains settling in the cold outer part of the PSN midplane \cite{Gu06} and that (ii) the disk experienced some chemical evolution in the giant planets formation region (loss of H$_2$ and He), due to photoevaporation. In this scenario, these icy grains migrated toward the formation region of the giant planet where they subsequently released their trapped noble gases, due to increasing temperature. Due to the disk's photoevaporation inducing fractionation between H$_2$, He and the other heavier species, these noble gases would have been supplied in supersolar proportions from the PSN gas to the forming giant planets. The other species, whose trapping/condensation temperatures are higher, would have been delivered to the envelopes of the giants in the form of amorphous ice or clathrates. \cite{Gu06} predict that, while supersolar, the noble gas enrichments should be more moderate than those resulting from the accretion of solids containing O, C, N, S by the two giants.

\item {\it CO Snowline Model} Another scenario, proposed by \cite{Al14}, suggests that Uranus and Neptune were both formed at the location of the CO snowline in a stationary disk. Due to the diffusive redistribution of vapors (the so-called {\it cold finger effect}; \cite{St88,Cy98}), this location of the PSN intrinsically had enough surface density to form both planets from carbon-- and oxygen--rich solids but nitrogen-depleted gas. The analysis has not been extended to the other volatiles but this scenario predicts that species whose snowlines are beyond that of CO remain in the gas phase and are significantly depleted in the envelope compared to carbon. Under those circumstances, one should expect that Ar presents the same depletion pattern as for N in the atmospheres of Uranus and Neptune. In contrast, Kr, Xe, S and P should be found supersolar in the envelopes of the two ice giants, but to a lower extent compared to the C and O abundances, which are similarly very high \cite{Al14}.

\end{itemize}

\paragraph{Summary of Key Measurements}
Here we list the key measurements to be performed by an atmospheric entry probe at Saturn, Uranus and Neptune to better constrain their formation and evolution scenarios:

\begin{itemize}[label=\textbullet]

\item Temperature--pressure profile from the stratosphere down to at least 10 bars. This would establish the stability of the atmosphere towards vertical motions and constrain the opacity properties of clouds lying at or above these levels (CH$_4$ and NH$_3$ or H$_2$S clouds). At certain pressures convection may be inhibited by the mean molecular weight gradient \cite{Gu95} (for instance at $\sim$2 bar in Neptune) and it is thus important to measure the temperature gradient in this region. Probing deeper than $\sim$40 bars would be needed to assess the bulk abundances of N and S existing in the form of NH$_4$SH but this would require microwave measurements from a Juno-like orbiter, instead of using a shallow probe.

\item Tropospheric abundances of C, N, S, and P, down to the 10-bar level at least, with accuracies of $\pm$10\%~(of the order of the protosolar abundance accuracies). In the case of the ice giants, N and S could be measured remotely deeper to the 40-bar level at microwave wavelengths by a Juno-like orbiter. 

\item Tropospheric abundances of noble gases He, Ne, Xe, Kr, Ar, and their isotopes to trace materials in the subreservoirs of the PSN. The accuracy on He should be at least as good that obtained by Galileo at Jupiter ($\pm$2\%), and the accuracy on isotopic ratios should be $\pm$1\% to enable direct comparison with other known Solar System values.

\item Isotopic ratios in hydrogen (D/H) and nitrogen ($^{15}$N/$^{14}$N), with accuracies of $\pm$5\%, and in oxygen ($^{17}$O/$^{16}$O and $^{18}$O/$^{16}$O) and carbon ($^{13}$C/$^{12}$C) with accuracies of $\pm$1\%. This will enable us to determine the main reservoirs of these species in the PSN.  

\item Tropospheric abundances of CO and PH$_3$. Having both values brackets the deep H$_2$O abundance \cite{Vi05}. CO alone may not be sufficient to enable the evaluation of the deep H$_2$O because of the uncertainties on the deep thermal profile (convection inhibition possible at the H$_2$O condensation level) as shown in \cite{Ca17}.

\end{itemize}

\begin{figure*}[t]
\centering 
\includegraphics[width=16.0cm]{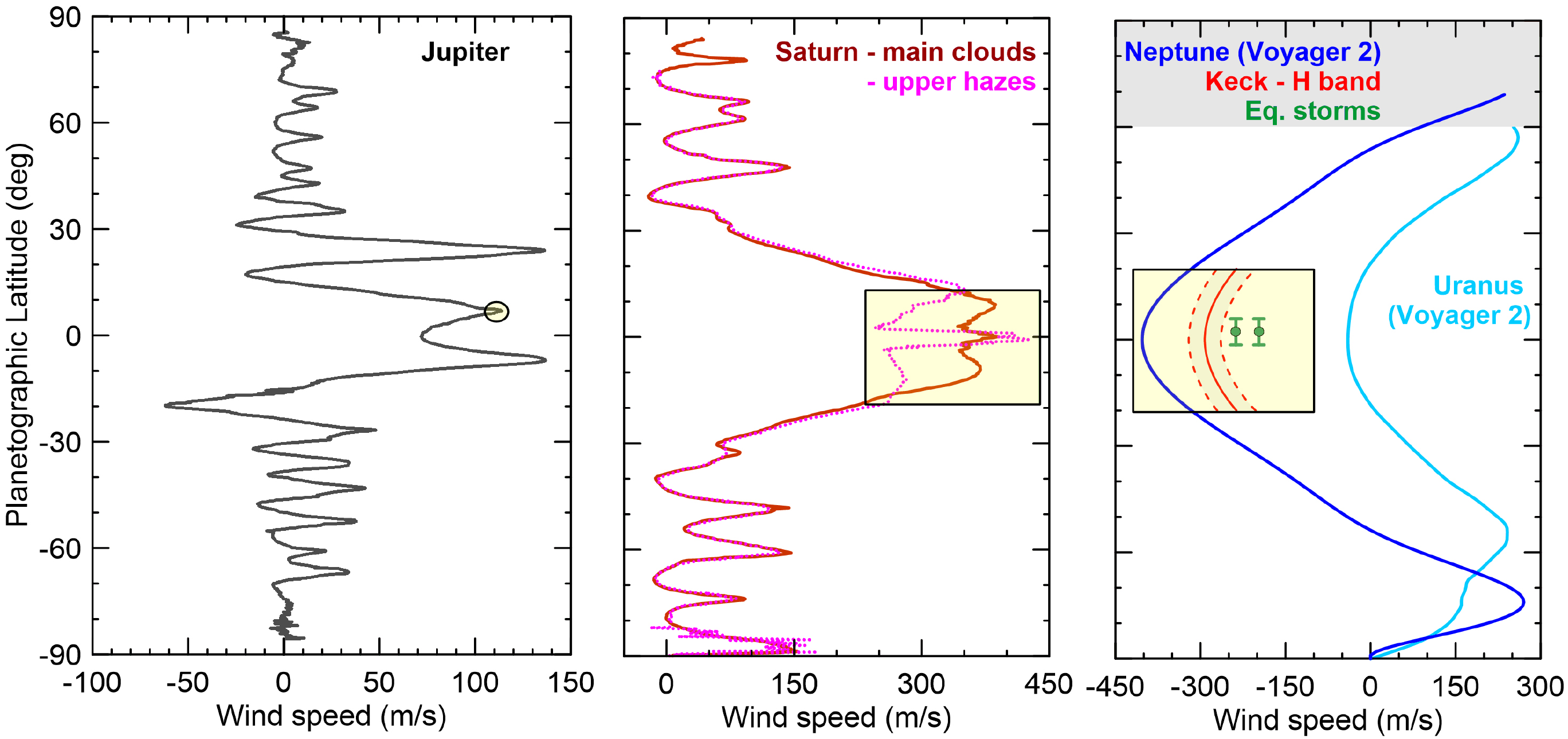}
\caption{Zonal winds in Jupiter, Saturn, Uranus and Neptune \cite{Sanchez-Lavega2019book} from different space missions. The equator, which is home to strong vertical wind-shear, is highlighted in each panel. In Jupiter the Galileo Probe measured strong vertical wind shears confined to the first 5 bar of the atmosphere \cite{Atkinson1996, Atkinson1997}. In Saturn \cite{Garcia-Melendo2011} and Neptune there are strong evidences of vertical wind shears at the equator \cite{Tollefson2018, Molter2019}.}
\label{fig:Winds}
\end{figure*}

\subsection{In Situ Studies of Giant Planet Atmospheres}

The giant planets are natural planetary-scale laboratories for the study of fluid dynamics without the complex effects of topography and ocean--atmosphere coupling. Remote sensing provides access to a limited range of altitudes, typically from the tropospheric clouds upwards to the lower stratosphere and thermosphere, although microwave radiation can probe deeper below the upper cloud deck. The vertical resolution of ``nadir'' remote sensing is limited to the width of the contribution function (i.e., the range of altitudes contributing to the upwelling radiance at a given wavelength), which can extend over one or more scale heights and makes it impossible to uniquely identify the temperature and density perturbations associated with cloud formation, wave phenomena, etc. {\it In situ} exploration of Saturn, Uranus or Neptune would not only constrain their bulk chemical composition, but it would also provide direct sampling and ``ground-truth'' for the myriad of physical and chemical processes at work in their atmospheres. In the following we explore the scientific potential for a probe investigating atmospheric dynamics, meteorology, clouds and hazes, and chemistry. We also provide the key atmospheric observables accessible to an atmospheric probe.

\paragraph{Zonal Winds} At the cloud tops, Jupiter and Saturn have multi-jet winds with eastward equatorial jets, while Uranus and Neptune have a broad retrograde equatorial jet and nearly symmetric prograde jets at high latitudes \cite{Sanchez-Lavega2019book} (Fig. \ref{fig:Winds}). The question of the origin of the jets and the differences between the gas giants and the icy giants is the subject of intensive research. Numerical attempts to study this question are based either on external forcing by the solar irradiation with a shallow circulation, or in deep forcing from the internal heat source of the planets producing internal columnar convection \cite{Vasavada2005, Sanchez-Lavega2019book}. However none of these models has been able to reproduce the characteristics of the wind systems of the planets without fine-tuning their multiple parameters.
It is possible to explore the depth of the winds through measurements of the gravity field of the planet combined with interior models. Recent results from Juno \cite{Kaspi2018, Guillot2018} and Cassini \cite{Galanti2019}, and a reanalysis of Uranus and Neptune Voyager data \cite{Kaspi2013} show that the winds are neither shallow, nor deep in any of these planets and may extend 3,000 km in Jupiter, 9,000 km in Saturn and 1,000 km in Uranus and Neptune. Vertical wind-shears are determined by measuring the horizontal distribution of temperature. Remote sensing can provide maps of temperature above the clouds but do not permit the determination of the deeper winds. In addition, in Uranus and Neptune, the horizontal distribution of volatiles causes humidity winds \cite{Sun1991}, an effect that occurs in hydrogen-helium atmospheres with highly enriched volatiles. 

{\it In situ} measurements of how the wind changes in the top few tens of bars (e.g., like Galileo) would provide insights into how the winds are being generated. The vertical wind shear measured by Galileo defied previous ideas of the expected structure of the winds. Theoretical models of atmospheric jets driven by solar heat flux and shallow atmospheric processes include a crucial role of moist convection in the troposphere \cite{Lian2010} and only through knowledge of the vertical distribution of condensables and winds we will be able to understand the generated wind systems of these planets.

\paragraph{Temperature Structure}
Vertical profiles of temperature in the upper atmospheres are retrieved from mid-infrared and sub-millimetre remote sounding. The determination of these vertical profiles from occultation measurements depends on the knowledge of the mean molecular weight, and therefore, requires simultaneous sensing of infrared radiance to constrain the bulk composition. However, measuring the vertical (and horizontal) distribution of volatile gases and their condensed phases from orbit is a fundamentally degenerate problem. Hence entry probes are the only way to determine these quantities with accuracy and provide a ground-truth to the study of the temperature distribution. This is true for Saturn even if the very successful Cassini mission has provided unprecedented observations of the temperature structure of the planet \cite{Fletcher2007}. Models of globally-averaged temperatures for Uranus \cite{Orton2014} and Neptune \cite{Fletcher2014} present  differences with the radio occultation results \cite{Lindal1987, Lindal1992} and an {\it in situ} determination of a thermal profile and vertical distribution of mean molecular weight is a vital measurement for the interpretation of thermal data. Furthermore, available data is limited to pressures smaller than 1 bar or is intrinsically degenerate and model-dependent. A considerable uncertainty in Uranus and Neptune is due to the molecular weight gradient caused by methane condensation and the resulting inhibition of moist convection in the atmosphere \cite{Gu95,Leconte2017,Friedson2017}, with a resulting temperature profile that may be sub-adiabatic, dry adiabatic or superadiabatic. This has consequences for interior and evolution models, atmospheric dynamics and the interpretation of abundances measurements in particular for disequilibrium species. In situ measurements will provide ground truth. Because in these planets the methane condensation region is at pressures smaller than 2 bars, this is well within reach of the probe that we consider. Also, solar irradiation alone cannot explain the high temperatures found in the stratospheres and thermospheres of Uranus and Neptune \cite{Herbert1987, Li2018}, a problem known as the energy crisis that cannot be solved from remote sensing. Measurements of temperatures in the stratosphere would result in a detailed characterization of gravity waves propagation that could help us to resolve energy transfer processes in planetary atmospheres in general.

\begin{figure}[t]
\centering 
\includegraphics[width=8.7cm]{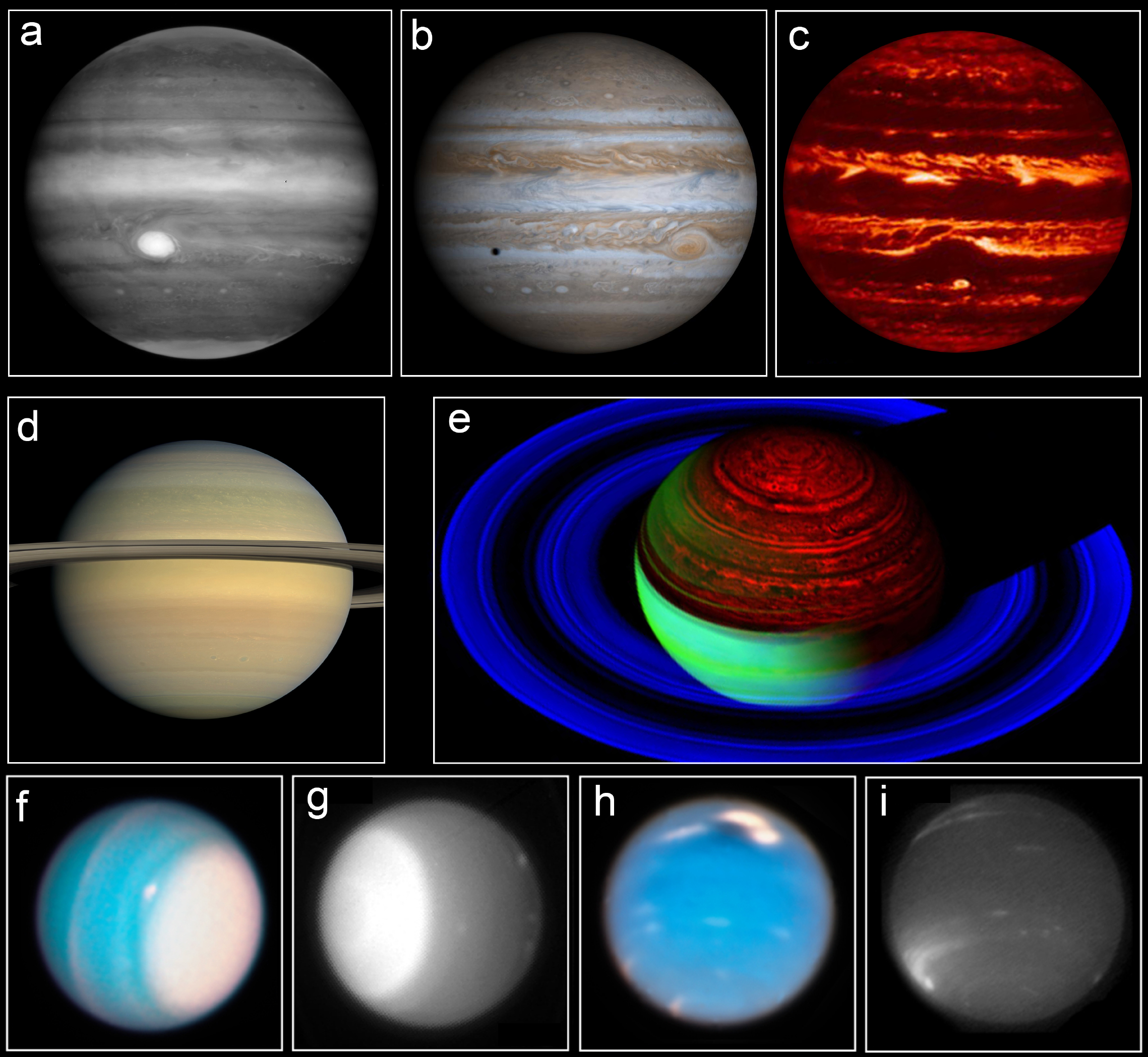}
\caption{Multi-wavelength images of Jupiter (upper row), Saturn (middle row) and Uranus and Neptune (bottom row). Images in the near-infrared in methane absorption bands (a, g, i) sample complex layers of hazes. Visible images (b, d, f, h) correspond to the top of the main upper cloud (NH$_3$ in Jupiter and Saturn and CH$_4$ in Uranus and Neptune). Infrared images at 4-5 $\mu$m(c, e) sample the opacity of a secondary cloud layer, most probably NH$_4$SH in Jupiter and Saturn.}
\label{fig:Images_spectra}
\end{figure}

\paragraph{Clouds} 
Images of the gas and ice giants in the visible and near-infrared show a plethora of clouds that organize in zonal bands, vortices, planetary waves and turbulent regions (Fig. \ref{fig:Images_spectra}). The vertical structure of clouds from multi-wavelength observations can be interpreted via radiative-transfer models, but these models offer multiple possibilities to fit individual observations and require a good knowledge of the vertical distribution of absorbing species like methane or volatile gases. The observable clouds in Jupiter and Saturn are separated in three layers (hazes close to the tropopause at 60--100 mbar, high-opacity clouds with their tops at 400--700 mbar and deep clouds with opacity sources at around 1.5--2.0 bar). The accessible clouds in Uranus and Neptune are different with an extended haze layer topping at 50--100 mbar located above a thin methane cloud of ice condensates with its base at $\sim1.3$~bar. This cloud is above another cloud of H$_2$S ice that is optically thick, located between 2 and 4 bar of pressure and whose structure can not be discerned from the observations. These basic vertical cloud structures come from multiple independent studies (\cite{Banfield1998,West2004,Perez-Hoyos2012} for Jupiter, \cite{West2009,Fletcher2011,Perez-Hoyos2016} for Saturn, \cite{West1991,Irwin2009,deKleer2015,Irwin2017} for Uranus, and \cite{Hammel1989,Baines1994,Irwin2016} for Neptune), and generally assume specific properties of the clouds in different regions of the planet. However, radiative transfer models produce highly degenerate solutions where multiple possibilities for the cloud particle optical properties and vertical structure can be found that can fit the observations. Under those circumstances, {\it in situ measurements provide a ground-truth to remote sensing observations. They give us information about clouds much deeper than what can be observed from remote sensing}.

\begin{figure}[t]
\centering 
\includegraphics[width=9cm]{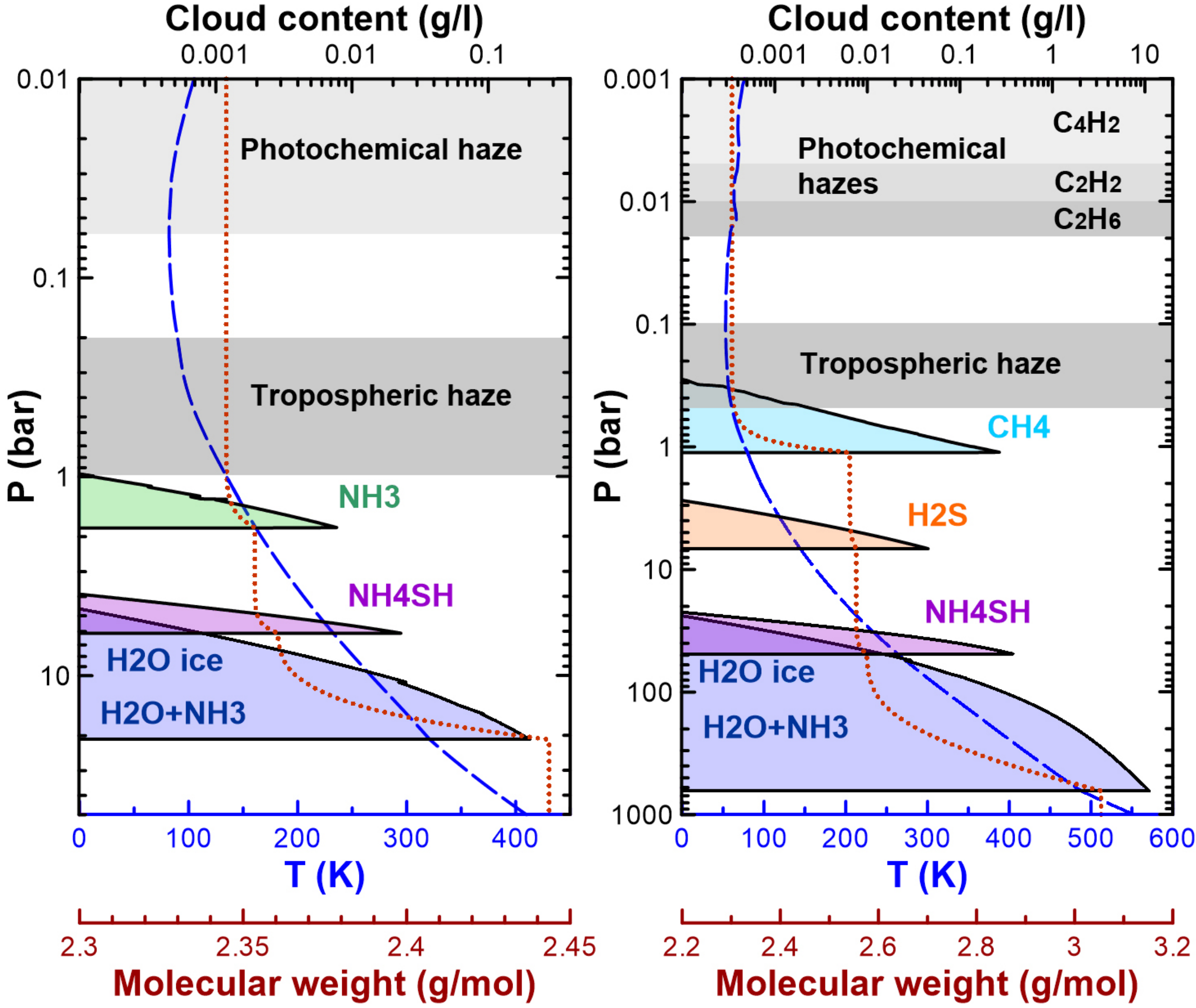}
\caption{Vertical cloud structure in Saturn (left) and Uranus (right) from Voyager thermal profiles extended following a moist adiabat (dashed-line) and assuming 5 times solar abundance of condensable for Saturn and 30 times solar abundances for Uranus except for NH$_3$, which is assumed to have a lower abundance than H$_2$S. The upper atmosphere is home to several photochemical layers. The vertical distribution of molecular weight is also shown (dotted-line). Simple ECC models do not take into account precipitation of condensates and the actual cloud structure could be different. Tropospheric hazes required by radiative-transfer models to fit the observations are also shown.}
\label{fig:cloud_models}
\end{figure}

The relation between the bands and colors in the giant planets is not well understood. The pattern of bands in Jupiter observed in the visible follows the structure of the zonal jets \cite{Ingersoll2004}. The same holds  partially in Saturn \cite{Vasavada2006}, but the bands in Uranus and Neptune have  a much richer structure than the wind field \cite{Smith1986,Smith1989,Sromovsky2001,Fry2012}. In all planets changes in the bands do not seem to imply changes in the more stable wind system \cite{Sanchez-Lavega2019book}. Questions about how the belt and circulation pattern can be established \cite{Fletcher2019} may require information from atmospheric layers below the visible pattern of clouds, which are not accessible to remote sensing. Exploring deeper into the atmosphere requires the use thermochemical Equilibrium Cloud Condensation (ECC) models which predict the location of clouds based on hypothesis of the relative abundances of condensables and thermal extrapolations of the upper temperatures \cite{Weidenschilling1973, Atreya2005}. Depending on the planet and relative abundances of the condensables several cloud layers are predicted to form: NH$_3$, NH$_4$SH and H$_2$O in Jupiter and Saturn, and CH$_4$, H$_2$S, NH$_4$SH and H$_2$O in Uranus and Neptune (Fig. \ref{fig:cloud_models}).  An additional intermediate cloud of NH$_3$ could form at pressures around 10 bar depending of the sequestration of NH$_3$ molecules in the lower NH$_4$SH cloud and the amount of NH$_3$ dissolved in the deep and massive liquid water cloud. This ammonia cloud is not expected currently in Uranus and Neptune due to the detection of tropospheric H$_2$S gas \cite{Irwin2018, Irwin2019} that seems to indicate that H$_2$S is more abundant than NH$_3$ in these atmospheres. 

A shallow probe to 10 bar in Saturn would descend below the NH$_4$SH cloud but may not probe the water cloud base and its deep abundance. A similar probe in Uranus and Neptune would descend below the H$_2$S cloud, while a deep probe would be needed to reach the NH$_4$SH cloud layer and the top of the H$_2$O cloud, which could extend to hundreds of bars. However, the descent profile would depend on the properties of the meteorological environment of the descent \cite{Showman1998}, a question we now examine.

\paragraph{Convection and Meteorological Features}   
Moist convection develops through the release of latent heat when gases condense and mix vertically impacting the vertical distribution of volatiles, molecular weight and temperature. In the giant planets volatiles are heavier than the dry air reducing the buoyancy of convective storms and potentially inhibiting moist convection in Jupiter's deep water cloud layer for water abundances higher than 5 and in Uranus and Neptune methane and deeper clouds \cite{Gu95, Leconte2017}. However, convective storms are relatively common in Jupiter and group in cyclonic regions \cite{Ingersoll2004, Vasavada2005}. In Saturn, they occur seasonally in the tropics over extended periods of time \cite{Fischer2019} and develop into Great Storms once per Saturn year \cite{Sanchez-Lavega2018_book}. Discrete cloud systems form and dissipate episodically in Uranus and Neptune including bright cloud systems that could be intense storms \cite{dePater2015, Molter2019}. However, there is no consensus whether or not these features are events of energetic moist convection as their vertical cloud structure does not result in the elevated cloud tops \cite{Irwin_Uranus2017} expected from comparison with Jupiter and Saturn and basic models of moist convection \cite{Stoker1989}. Large and small vortices, waves and turbulent regions are common in the atmospheres of Jupiter and Saturn \cite{Ingersoll2004, Showman2018}. Neptune is famous for its dark vortices surrounded by bright companion clouds \cite{Smith1989, Wong2018} and Uranus has rare dark vortices \cite{Hammel2009} and bright cloud systems \cite{Sromovsky2015}. Many of these meteorological systems last for years to decades but we ignore how deep they extend into the lower troposphere. Large-scale waves can also affect the properties of the atmosphere well below the upper cloud layer \cite{Showman1998}.

{\it The interpretation of vertical profiles of pressure, temperature, wind speed, and composition obtained by a probe would hugely benefit from an observational characterization of the descending region and its meteorology at cloud level \cite{Orton1998}}.

\paragraph{Chemistry}
In the upper atmospheres of Jupiter, Saturn, Uranus and Neptune, methane is photolysed into hydrocarbons that diffuse down and condense to form haze layers in the cold stratospheres (altitudes $\sim$0.1 to 30 mbar) as the temperature decreases down to $\sim$60 K in the tropopause in Uranus and Neptune. Photochemical models suggest hazes made of hydrocarbons that become progressively more important from Jupiter to Uranus and Neptune with C$_2$H$_2$, C$_6$H$_6$, C$_4$H$_2$, C$_4$H$_{10}$, CO$_2$, C$_3$H$_8$, C$_2$H$_2$, add C$_2$H$_6$ \cite{Taylor2004, Fouchet2009, Moses2018}, where the oxygen species derive from external sources such as interplanetary dust or comets. These  species are radiatively active at mid-infrared wavelengths and affect the aerosol structure and  energy balance of the atmospheres and, thus, their overall dynamics. Tropospheric CO is particularly important because it is related with other oxygen bearing molecules including water. Thermochemical models have been used to relate the observed CO abundance with the deep water abundance \cite{Ca17} but results of these models depend on precise measurements of tropospheric CH$_4$ and knowledge of vertical mixing that can only be determined precisely {\it in situ}.

\paragraph{Summary of Key Measurements}
Below are indicated the key {\it in situ} measurements needed to characterize the atmospheres of Saturn, Uranus or Neptune.

\begin{itemize}[label=\textbullet]

\item Temperature-pressure profile. This basic but essential measurement will be key to check widespread but model-dependent measurements obtained from remote observations. Testing for the presence of sub- or super-adiabatic lapse rates will be key to understand how internal heat is transported in these active atmospheres. 

\item Cloud and haze properties. A descent probe would be able to measure the atmospheric aerosols scattering properties at a range of phase angles, the particles number density, the aerosol shape and opacity properties. Each of these measurements would help constrain the aerosol composition, size, shape, and density.

\item Winds. Doppler wind measurements provide the wind profile in the lower troposphere, well below  the region where most of the cloud tracking wind measurements are obtained. Static and dynamic pressures would provide an estimate of the vertical winds, waves, and convection. The comparison with vertical profiles of condensable abundances and thermal data would quantify the relative importance of thermal and humidity winds.

\item Conductivity.  A vertical profile of atmospheric conductivity would indicate what type of clouds support charge separation to generate lightning. Conductivity measurements combined with meteorological and chemical data (particularly measurements of the physical properties of the aerosols themselves) would also permit extraction of the charge distribution on aerosol particles, and improve understanding of the role of electrical processes in cloud formation, lightning generation, and aerosol microphysics. 

\item Determine the influence of cloud condensation or photochemical haze formation on the temperature lapse rate and deduce the amount of energy relinquished by this phase change in key species (CH$_4$, NH$_3$, H$_2$S).

\item Ortho-to-para hydrogen ratio. This would constrain the degree of vertical convection through the atmosphere and the convective capability at different cloud condensing layers. It would also be essential to understand the vertical profile of atmospheric stability and is especially important in the cold atmospheres of Uranus and Neptune.

\end{itemize}

\section{Mission Configuration and Profile}


\subsection{Probe Mission Concept}

The three giant planets considered in this White Paper can be targeted with a similar probe payload and architecture.

\paragraph{Science Mission Profile} 

To measure the atmospheric composition, thermal and energy structure, clouds and dynamics requires {\it in situ} measurements by a probe carrying a mass spectrometer (atmospheric and cloud compositions), helium abundance detector, atmospheric structure instrument (thermal structure and atmospheric stability), nephelometer (cloud locations and aerosol properties), net flux radiometer (energy structure), physical properties instrument (temperature, pressure and density structure, ortho-para ratio), and Doppler-wind experiment (dynamics). The atmospheric probe descent targets the 10-bar level located about 5 scale heights beneath the tropopause. The speed of probe descent will be affected by requirements imposed by the needed sampling periods of the instruments, particularly the mass spectrometer, as well as the effect speed has on the measurements. This is potentially an issue for composition instruments, and will affect the altitude resolution of the Doppler wind measurement. Although it is expected that the probe batteries, structure, thermal control, and telecomm will allow operations to levels well below 10 bars, a delicate balance must be found between the total science data volume requirements to achieve the high-priority mission goals, the capability of the telecomm system to transmit the entire science, engineering, and housekeeping data set (including entry accelerometry and pre-entry/entry calibration, which must be transmitted interleaved with descent data) within the descent telecomm/operational time window, and the probe descent architecture which allows the probe to reach 10 bars, i.e. {\it the depth at which most of the science goals can be achieved}.

\paragraph{Probe Mission Profile to Achieve Science Goals} A giant planet probe designed for parachute descent to make atmospheric measurements of composition, structure, and dynamics, with data returned to Earth using an orbiting or flyby Carrier Relay Spacecraft (CRSC) could be carried as an element of a dedicated giant planet system exploration mission. The CRSC would receive and store probe science data in real-time, then re-transmit the science and engineering data to Earth. While recording entry and descent science and engineering data returned by the probe, the CRSC would additionally make measurements of probe relay link signal strength and Doppler for descent probe radio science. Carried by the CRSC into the vicinity of the giant planet system, the probe would be configured for release, coast, entry, and atmospheric descent. For proper probe delivery to the entry interface point, the CRSC with probe attached is placed on a planetary entry trajectory, and is reoriented for probe targeting and release. The probe coast timer and pre-programmed probe descent science sequence are loaded prior to release from the CRSC, and following spin-up, the probe is released for a ballistic coast to the entry point. Following probe release, a deflect maneuver is performed to place the CRSC on the proper overflight trajectory to receive the probe descent telemetry. 

Prior to arrival at the entry interface point, the probe coast timer awakens the probe for sequential power-on, warm-up, and health checks. The only instrumentation collecting data during entry would be the entry accelerometers and possibly heat shield instrumentation including ablation sensors. The end of entry is determined by the accelerometers, initiating parachute deployment, aeroshell release, and the probe atmospheric descent. Parachute sequence would be initiated above the tropopause by deploying a pilot parachute which pulls off the probe aft cover, thereby extracting the main descent parachute, followed by release of the probe heatshield and initiation of a transmit-only telecommunications link from the probe to the CRSC. Under the parachute, the altitude of any required descent science operation mode changes would be guided by input from the Atmospheric Structure Instrument sensors, thereby providing the opportunity to optimize the data collection for changing science objectives at different atmospheric depths. The probe science data collection and relay transmission strategy would be designed to ensure the entire probe science data set is successfully transmitted prior to probe reaching the targeted depth. 

\begin{figure*}[ht]
\centering 
\includegraphics[width=12cm]{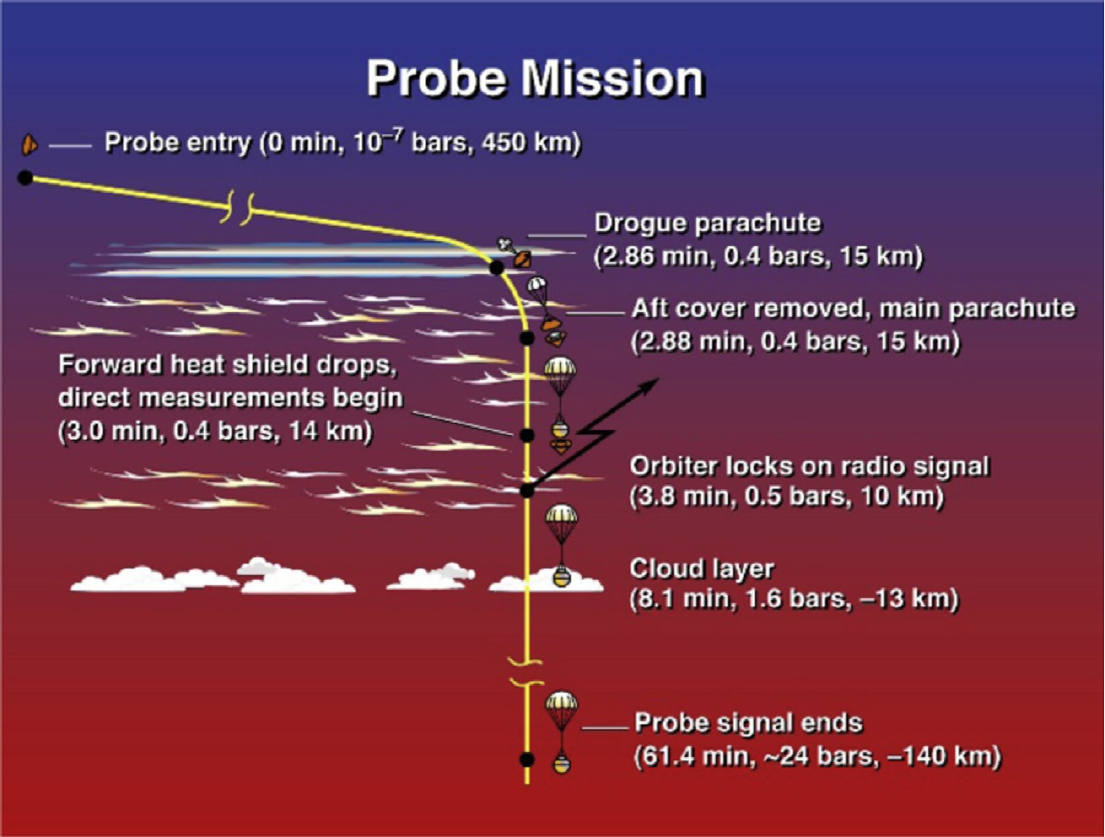}
\caption{Galileo entry, descent and deployment sequence shown above could be the basis for any proposed giant planet entry probe mission.}
\label{fig:entry}
\end{figure*}

Probe descent mission would likely end when the telecomm geometry becomes so poor that the link can no longer be maintained due to increasing overhead atmospheric opacity, depletion of the batteries, or increasing and damaging thermal and/or pressure effects. The probe transmits science and engineering data to the CRSC where multiple copies are stored in redundant on-board memory. At the completion of the probe descent mission and once the post-descent context observations have been performed, the CRSC reorients to point the High Gain Antenna towards Earth and all stored copies of the probe science and engineering data are returned to Earth. Figure \ref{fig:entry} represents a schematic view of the Galileo entry, descent and deployment sequence which could be the basis for any proposed entry probe mission.

\subsection{Probe Delivery}
\paragraph{Interplanetary Trajectory} Four characteristics of interplanetary transfers from Earth to the giant planets are of primary importance: 1) the launch energy affecting the delivered mass, 2) the flight time which affects required spacecraft reliability engineering and radioisotope power systems whose output power decreases with time, 3) the $V_\infty$ of approach (VAP) to the destination planet which influences the $\Delta$$V$ necessary for orbit insertion and the entry speed of an entry probe delivered from approach, and 4) the declination of the approach (DAP) asymptote which influences both the locations available to an entry probe and the probe's atmosphere-relative entry speed which depends on the alignment of the entry velocity vector with the local planetary rotation velocity. 
Depending on transfer design and mass, trajectories to the giant planets can be order of 5--6 years for Jupiter, up to 10--13 years for Uranus and Neptune. When Jupiter and Saturn align to provide gravity assists from both, trajectories with shorter transfer durations are possible. 

\paragraph{Probe Delivery and Options for Probe Entry Location} Gi- ven a transfer trajectory defined by its VAP and DAP, a remaining degree of freedom - the ``b'' parameter (the offset of the b-plane aim point from the planet's center), determines both the available entry site locations, and the atmosphere-relative entry speed for each of those locations, and the entry flight path angle (EFPA). If the probe is delivered and supported by a flyby spacecraft, designing a trajectory to give data relay window durations of an hour or more is not difficult. However, if the CRSC is an orbiter delivering the probe from hyperbolic approach, the probe mission must compete with the orbit insertion maneuver for best performance. Although orbit insertion maneuvers are most efficiently done near the pMi17lanet thereby saving propellant mass, such trajectories coupled with a moderately shallow probe EFPA that keeps entry heating rates and inertial loads relatively low would yield impractically short data relay durations. For the ice giants, a different approach to this problem might avoid this situation by delivering the probe to an aim point $\sim$180$\deg$ away from the orbiter's aim point. Although this requires a minor increase in the orbiter's total $\Delta$$V$ for targeting and deflection, it allows a moderate EFPA for the probe while providing a data relay window of up to 2 hours.

\paragraph{Probe Entry and Enabling Technologies} The probe aer- oshell would comprise both a forward aeroshell (heatshield) and an aft cover (backshell). The aeroshell has five primary functions -- 1) to provide an aerodynamically stable configuration during hypersonic and supersonic entry and descent into the giant planet H$_2$--He atmosphere while spin-stabilized along the probe's symmetry (rotation) axis, 2) to protect the descent vehicle from the extreme heating and thermomechanical loads of entry, 3) to accommodate the large deceleration loads from the descent vehicle during hypersonic entry, 4) to provide a safe, stable transition from hypersonic/supersonic entry to subsonic descent, and 5) to safely separate the heatshield and backshell from the descent vehicle based on g-switch with timer backup, and transition the descent vehicle to descent science mode beneath the main parachute. The need for a heatshield to withstand the extreme entry conditions encountered at the giant planets is critical and has been successfully addressed by NASA in the past, and is currently addressed by ESA. Because heritage carbon phenolic thermal protection system (TPS) used for the Galileo and Pioneer Venus entry aeroshell heatshields is no longer available, NASA invested in the development of a new heatshield material and system technology called Heatshield for Extreme Entry Environment Technology (HEEET) and also in upgrading arc jet facilities for ablative material testing at extreme conditions. HEEET is an ablative TPS system that uses 3-D weaving to achieve both robustness and mass efficiency at extreme entry conditions, and being tested at conditions that are relevant for Saturn and ice giant entry probe missions \cite{Vl16}. Compared to heritage carbon phenolic system, HEEET is nearly 50\% mass efficient \cite{Mi17}. Alternative TPS concepts and materials are currently under evaluation by ESA (ESA M* Ice Giant CDF study 2).

\subsection{Atmospheric Entry Probe System Design}
\paragraph{Overview} The probe comprises two major sub-elements: 1) the Descent Vehicle (DV) including parachutes will carry all the science instruments and support subsystems including telecommunications, power, control, and thermal into the atmosphere, and 2) the aeroshell that protects the DV during cruise, coast, and entry. The probe (DV and aeroshell) is released from the CRSC, and arrives at the entry interface point following a long coast period. Although the probe reaches the entry interface point and the DV with parachutes descends into the atmosphere, elements of the probe system including the probe release and separation mechanism and the probe telemetry receiver remain with the CRSC. Prior to entry, the probe coast timer (loaded prior to probe release) provides a wakeup call to initiate the entry power-on sequence for initial warmup, checks on instrument and subsystem health and status, and pre-entry calibrations. Entry peak heating, total heat soak, and deceleration pulse depend on the selected mission design including entry location (latitude/longitude), inertial heading, and flight path angle. Following entry, the DV provides a thermally protected environment for the science instruments and probe subsystems during atmospheric descent, including power, operational command, timing, and control, and reliable telecommunications for returning probe science and engineering data. The probe avionics will collect, buffer, format, process (as necessary), and prepare all science and engineering data to be transmitted to the CRSC. The probe descent subsystem controls the probe descent rate and rotation necessary to achieve the mission science objectives.

\paragraph{Entry Probe Power and Thermal Control} Following release from the CRSC, the probe has four main functions: 1) to initiate the ``wake up'' sequence at the proper time prior to arrival at the entry interface point, 2) to safely house, protect, provide command and control authority for, provide power for, and maintain a safe thermal environment for all the subsystems and science instruments, 3) to collect, buffer as needed, and relay to the CRSC all required preentry, entry, and descent housekeeping, engineering, calibration, and science engineering data, and 4) to control the descent speed and spin rate profile of the descent vehicle to satisfy science objectives and operational requirements. Once released from the CRSC, the probe would be entirely self-sufficient for mission operations, thermal control, and power management. During coast, pre-entry, and entry, the batteries support probe coast functions, wake-up and turn-on, system health checks, and entry and descent operations. Autonomous thermal control is provided during coast by batteries, although there may be an option to replace electrical heating with Radioisotope Heater Units to greatly reduce battery requirements. Giant planet missions may include Venus flybys, where temperatures are higher, prior to the long outer solar system cruise. Since the ice giants are much cooler than the gas giants, descent survival at the low ice giant temperatures may dictate a sealed probe. Providing a safe, stable thermal environment for probe subsystems and instruments over this range of heliocentric distances will require careful thermal design. Future technology developments may realize batteries with higher specific energies resulting in potential mass savings, and the development of electronics operating at cryogenic temperatures. 

\paragraph{Data Relay} The transmit-only probe telecommunication system would comprise two redundant channels that transmit orthogonal polarizations at slightly offset frequencies for isolation. Driven by an ultrastable oscillator to ensure a stable link frequency for probe radio science, the frequency of the probe to CRSC relay link is chosen primarily based on the microwave absorption properties of the atmosphere. The actual thermal, compositional, and dynamical structure beneath the cloud tops of the giant planets remains largely unknown. Possible differences in composition and temperature and pressure structure between the atmosphere models and the true atmosphere may adversely affect the performance of the probe relay telecomm and must be considered in selection of communication link frequency. In particular, the microwave opacity of the atmosphere depends on the abundance of trace microwave absorbing species such as H$_2$O, NH$_3$, H$_2$S, and PH$_3$. In general, the microwave opacity of these absorbers increases as the square of the frequency, and this drives the telecomm frequency as low as reasonable, often UHF. At Jupiter, the lowest practical frequency is L-band due to the intense low-frequency synchrotron radiation environment. The final decision on frequency consequently affects the overall telecomm link budget, including probe transmit antenna design (type, size, gain, and beam pattern, and beam-width), and pointing requirements for the CRSC-mounted receive antenna. Other decisions affecting the telecomm link design include probe descent science requirements, the time required to reach the target depth, and the CRSC overflight trajectory, including range, range rate, and angle.

\paragraph{Carrier Relay Spacecraft} During the long cruise to the outer solar system, the CRSC provides power as well as structural and thermal support for the probe, and supports periodic health checks, communications for probe science instrument software changes and calibrations, and other probe power and thermal control software configuration changes and mission sequence loading as might be required from launch to encounter. Upon final approach, the CRSC supports a final probe health and configuration check, rotates to the probe release orientation, cuts cables and releases the probe for the probe cruise to the entry interface point. Following probe release, the CRSC may be tracked for a period of time from Earth, preferably several days, to characterize the probe release dynamics and improve reconstructions of the probe coast trajectory and entry interface location. An important release sequence option would be to image the probe following release for optical navigation characterization of the release trajectory. Following probe release and once the CRSC tracking period is over, the CRSC is deflected from the planet-impact trajectory required for probe targeting to a trajectory that will properly position the CRSC for receiving the probe descent telecommunications. During coast, the probe will periodically transmit health status reports to the CRSC. Additionally, the CRSC will conduct a planet-imaging campaign to characterize the time evolution of the atmosphere, weather, and clouds at the probe entry site, as well as to provide global context of the entry site. Prior to the initiation of the probe descent sequence, the CRSC will rotate to the attitude required for the probe relay receive antenna to view the probe entry/descent location and subsequently prepares to receive both channels of the probe science telecommunications. Once the probe science mission ends, the CRSC will return to Earth-point and downlink multiple copies of the stored probe data.

\section{Possible Probe Model Payload}

Table \ref{tab:payload} presents a suite of scientific instruments that can address the scientific requirements discussed in Section \ref{science}. This list of instruments should be considered as an example of scientific payload that one might wish to see onboard. Ultimately, the payload of a giant planet probe would be defined from a detailed mass, power and design trades, but should seek to address the majority of the scientific goals outlined in Section \ref{science}.

\begin{table*}[h]
\begin{center}
\caption[]{Measurement requirements}
\small{\begin{tabular}{ll}
\toprule
Instrument					& Measurement									\\
\midrule
Atmospheric Structure Instrument	& Pressure, temperature, density, molecular weight profile	,	\\
							& atmospheric conductivity, DC electric field				\\
 	
Mass spectrometer				& Elemental and chemical composition					\\
							& Isotopic composition								\\
							& High molecular mass organics						\\
Tunable Laser System			& Isotopic composition								\\						
Helium Abundance Detector		& Helium abundance									\\	
Ortho-Para Instrument			& Temperature, pressure and density vertical structure		\\					
Doppler Wind Experiment			& Measure winds, speed and direction					\\						
Nephelometer					& Cloud structure									\\
							& Solid/liquid particles								\\
Net-Flux Radiometer				& Thermal/solar energy								\\									
\bottomrule
\end{tabular}}
\label{tab:payload}
\end{center}
\end{table*}

\paragraph{Atmospheric Structure Instrument} The Atmospheric Structure Instrument (ASI) is a multi-sensor package for {\it in situ} measurements to investigate the atmospheric structure, dynamics and electricity of the outer planets. The scientific objectives of ASI are the determination of the atmospheric vertical pressure and temperature profiles, the evaluation of the density, and the investigation of the atmospheric electrical properties (e.g. conductivity, lightning). The atmospheric profiles along the entry probe trajectory will be measured from the exosphere down deep into the outer planet's atmosphere. During entry, density will be derived from the probe decelerations; pressure and temperature will be computed from the density with the assumption of hydrostatic equilibrium. Direct measurements of pressure, temperature and electrical properties will be performed under the parachute, after the front shield jettisoning, by sensors having access to the atmospheric flow. ASI will measure the atmospheric state (pressure, temperature) as well as constraining atmospheric stability, dynamics and its effect on atmospheric chemistry. The ASI benefits from the strong heritage of the Huygens HASI experiment of the Cassini/Huygens mission \cite{Fu02}, and the Galileo and Pioneer Venus ASI instruments \cite{Se92,Se80}. 

\begin{figure}[ht]
\centering 
\includegraphics[width=9cm]{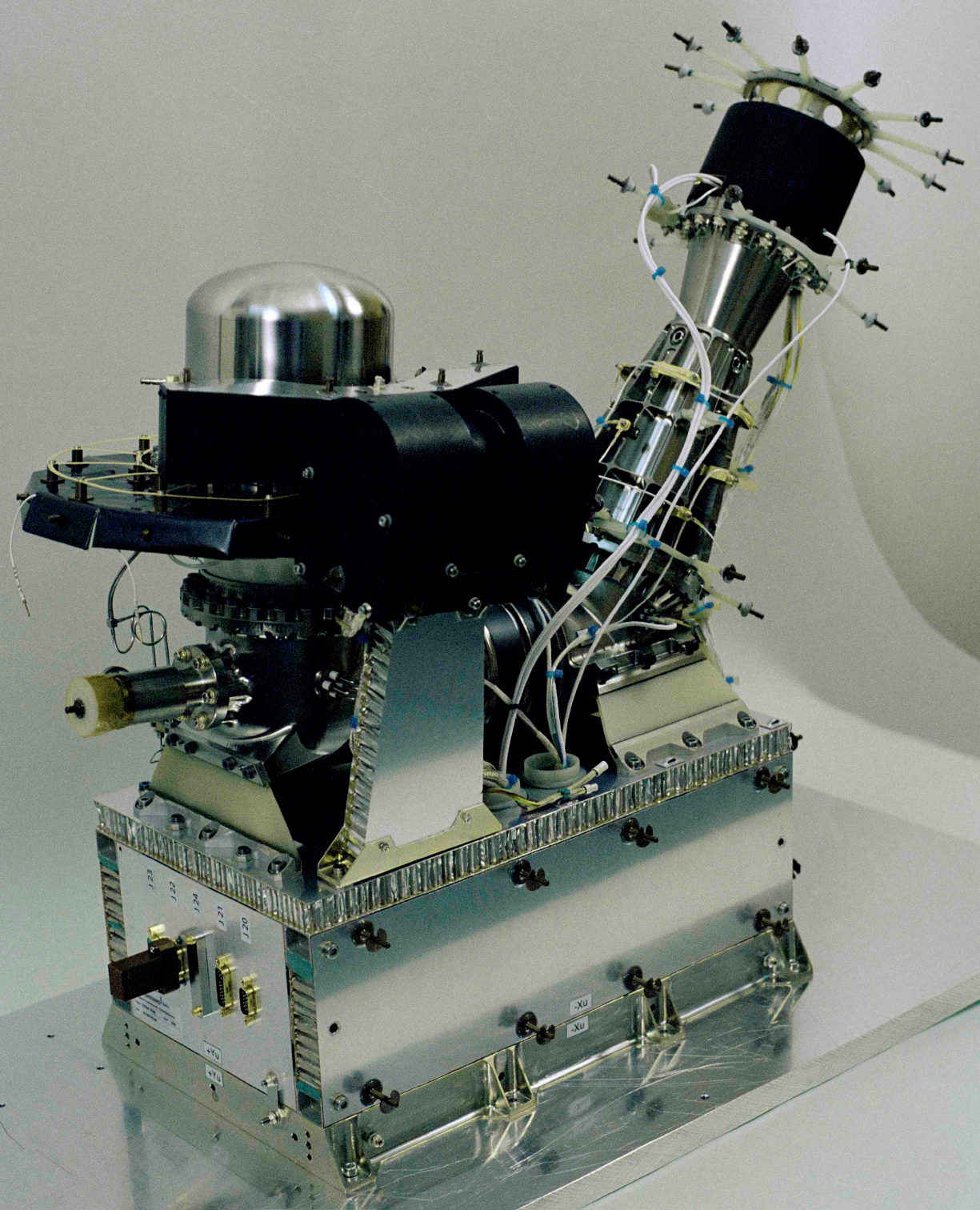}
\caption{Flight model of DFMS/ROSINA instrument without thermal hardware \cite{Ba07}.}
\label{fig:MS}
\end{figure}

\paragraph{Mass Spectrometer Experiment} 
The Mass Spectrometer Experiment (MSE) of the entry probe makes {\it in situ} measurements during the descent into the giant planets atmospheres to determine the chemical and isotopic composition of Uranus and Neptune. The scientific objective of MSE is to measure the chemical composition of  the major atmospheric species such as H, C, N, S, P, Ge, and As, all the noble gases He, Ne, Ar, Kr, and Xe, and key isotope ratios of major elements D/H, $^{13}$C/$^{12}$C, $^{15}$N/$^{14}$N, $^{17}$O/$^{16}$O, $^{18}$O/$^{16}$O, of the lighter noble gases $^{3}$He/$^{4}$He, $^{20}$Ne/$^{22}$Ne, $^{38}$Ar/$^{36}$Ar, $^{36}$Ar/$^{40}$Ar, and those of Kr and Xe. Given the constrained resources on the entry probe and the short duration of the descent through the atmosphere, time-of-flight instruments are the preferred choice, with strong heritage from the ROSINA experiment on the Rosetta mission \cite{Ba07} (see Fig. \ref{fig:MS}). The mass spectrometer itself will be complemented by a complex gas introduction system handling the range of atmospheric pressures during descent, a reference gas calibration system, and enrichment cells for improving the detection of noble gases and hydro carbons. 

\paragraph{Tunable Laser Spectrometer}
A Tunable Laser Spectrometer (TLS) \cite{Du10} will complement the mass spectrometric measurements by providing a few isotopic measurements with high accuracy, e.g. D/H, $^{13}$C/$^{12}$C, $^{18}$O/$^{16}$O, and $^{17}$O/$^{16}$O, depending on the selected laser system. TLS employs ultra-high spectral resolution (0.0005 cm$^{-1}$) tunable laser absorption spectroscopy in the near infra-red (IR) to mid-IR spectral region. A TLS is part of the SAM instrument on the NASA Curiosity Rover \cite{We11}, which was used to measure the isotopic ratios of D/H and of $^{18}$O/$^{16}$O in water and $^{13}$C/$^{12}$C, $^{18}$O/$^{16}$O, $^{17}$O/$^{16}$O, and $^{13}$C$^{18}$O/$^{12}$C$^{16}$O in carbon dioxide in the Martian atmosphere \cite{We13}.

\begin{figure}[ht]
\centering 
\includegraphics[width=9cm]{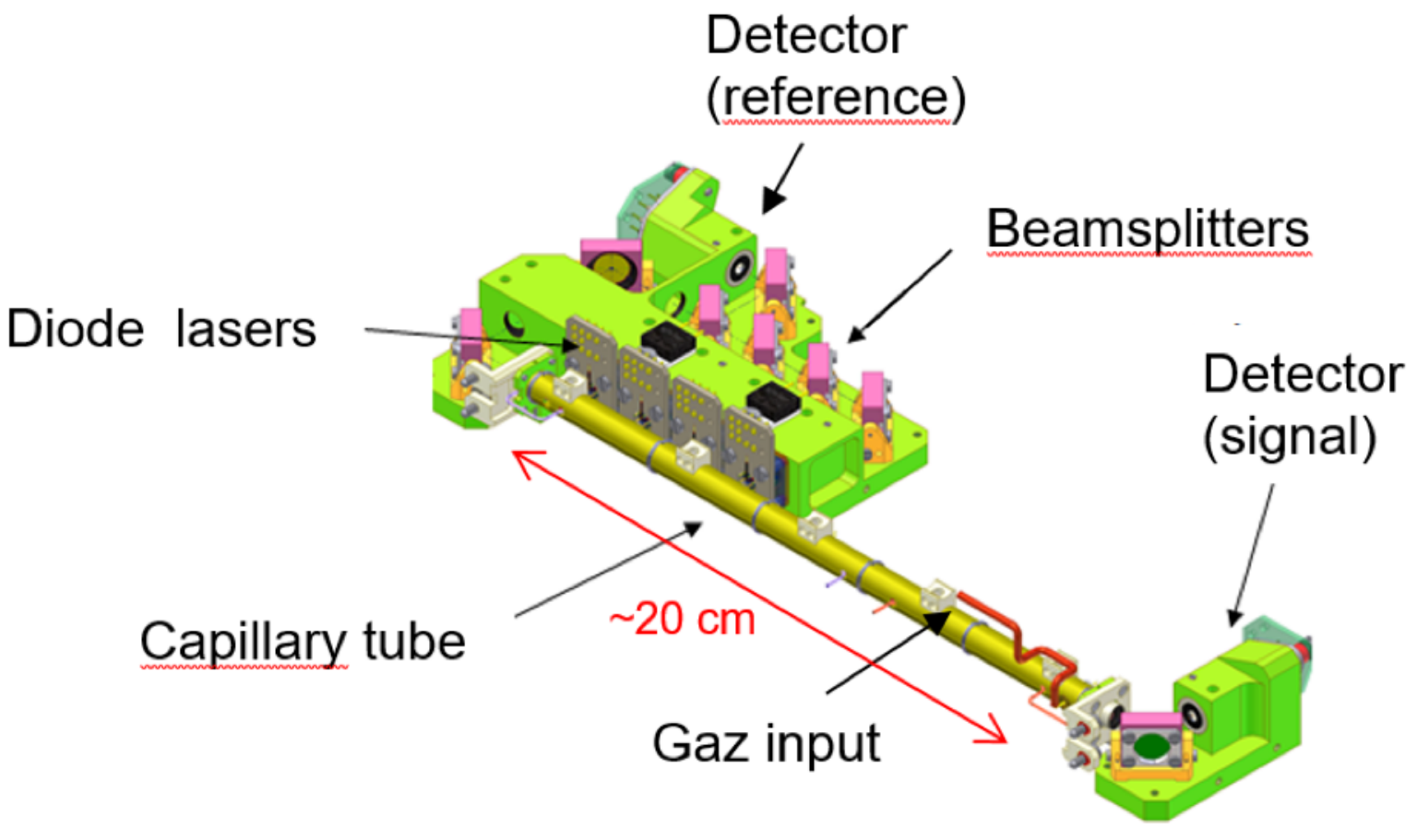}
\caption{A schematics of the laboratory model of the TLS spectrometer for the Martian Phobos Grunt mission \cite{Du10}. With the TLS, four near-infrared laser diodes are injected in a single-path tube filled up with the gases to analyse. The laser beam are partially absorbed by the ambient molecules. The gas concentrations for the various isotopologues are then retrieved from the achieved absorption spectra.}
\label{fig:TLS}
\end{figure}

\paragraph{Helium Abundance Detector}

The Helium Abundance Detector (HAD), as it was used on the Galileo mission \cite{vo92,vo98}, measures the refractive index of the atmosphere in the pressure range of 2--10 bar. The refractive index is a function of the composition of the sampled gas, and since the jovian atmosphere consists of mostly of H$_2$ and He, to more than 99.5\%, the refractive index is a direct measure of the He/H$_2$ ratio. The refractive index can be measured by any two-beam interferometer, where one beam passes through a reference gas and the other beam through atmospheric gas. The difference in the optical path gives the difference in refractive index between the reference and atmospheric gas. For the Galileo mission, a Jamin-Mascart interferometer was used, because of its simple and compact design, with a high accuracy of the He/H$_2$ measurement.

\paragraph{Doppler-Wind Experiment} The Doppler Wind Experiment (DWE) will use the probe-CRSC radio subsystem (with elements mounted on both the probe and the Carrier) to measure the altitude profile of zonal winds along the probe descent path under the assumption that the probe in terminal descent beneath the parachute will move with the winds. The DWE will also reflect probe motions due to atmospheric turbulence, aerodynamic buffeting, and atmospheric convection and waves that disrupt the probe descent speed. Key to the Doppler wind measurement is an accurate knowledge of the reconstructed probe location at the beginning of descent, the probe descent speed with respect to time/altitude, and the CRSC position and velocity throughout the period of the relay link. The initial probe descent location depends upon the probe entry trajectory from the entry point to the location of parachute deployment and is reconstructed from measured accelerations during entry. The descent profile is reconstructed from Atmospheric Structure Instrument measurements of pressure and temperature during descent. From the reconstructed probe and CRSC positions and velocities, a profile of the expected relay link frequencies is found that can be differenced with the measured frequencies to generate a set of frequency residuals. The winds are retrieved utilizing an inversion algorithm similar to the Galileo probe Doppler Wind measurement \cite{At97,At98}. To generate the stable probe relay signal, the probe must carry an ultrastable oscillator (USO) with an identical USO in the relay receiver on the Carrier spacecraft.

\paragraph{Nephelometer} Measurement of scattered visible light within the atmosphere is a powerful tool to retrieve number density and size distribution of liquid and solid particles, related to their formation process, and to understand the overall character of the atmospheric aerosols based on their refractive index (liquid particles, iced particles, solid particles from transparent to strongly absorbing). In particular, measurements of light scattered by a cloud of particles at several scattering angles was already tested on balloon flights to characterize the atmospheric aerosols and condensates \cite{Ga97}, using {\it a priori} hypothesis on the size distribution. A new concept of nephelometer has been proposed to retrieve the full scattering function, this time for individual particles crossing a light source. Dedicated fast electronics are necessary to enable the detection of up to 1,000 particles per cm$^3$. Such an instrument performs counting measurements at a small scattering angle, to retrieve the size distribution based on the work of \cite{Lu14}. It applies the principle of the Light Optical Aerosol Counter (LOAC) optical aerosols counter used since 2013 under all kinds of atmospheric balloons \cite{Re16a,Re16b}. These measurements allow one to retrieve the size distribution of the particles typically for 20 size-classes in the 0.2-50 $\mu$m range. Also, simultaneous measurements can be conducted at up to 10 scattering angles in the 20--170$^{\circ}$ range, to retrieve the scattering function for each size range. The retrieval of the nature of the aerosols can be conducted by comparing these observed scattering functions to theoretical ones computed for scattering theories, and to reference measurements obtained in laboratory for solid particles \cite{Re02,Vo06}.

\begin{figure}[ht]
\centering 
\includegraphics[width=8cm]{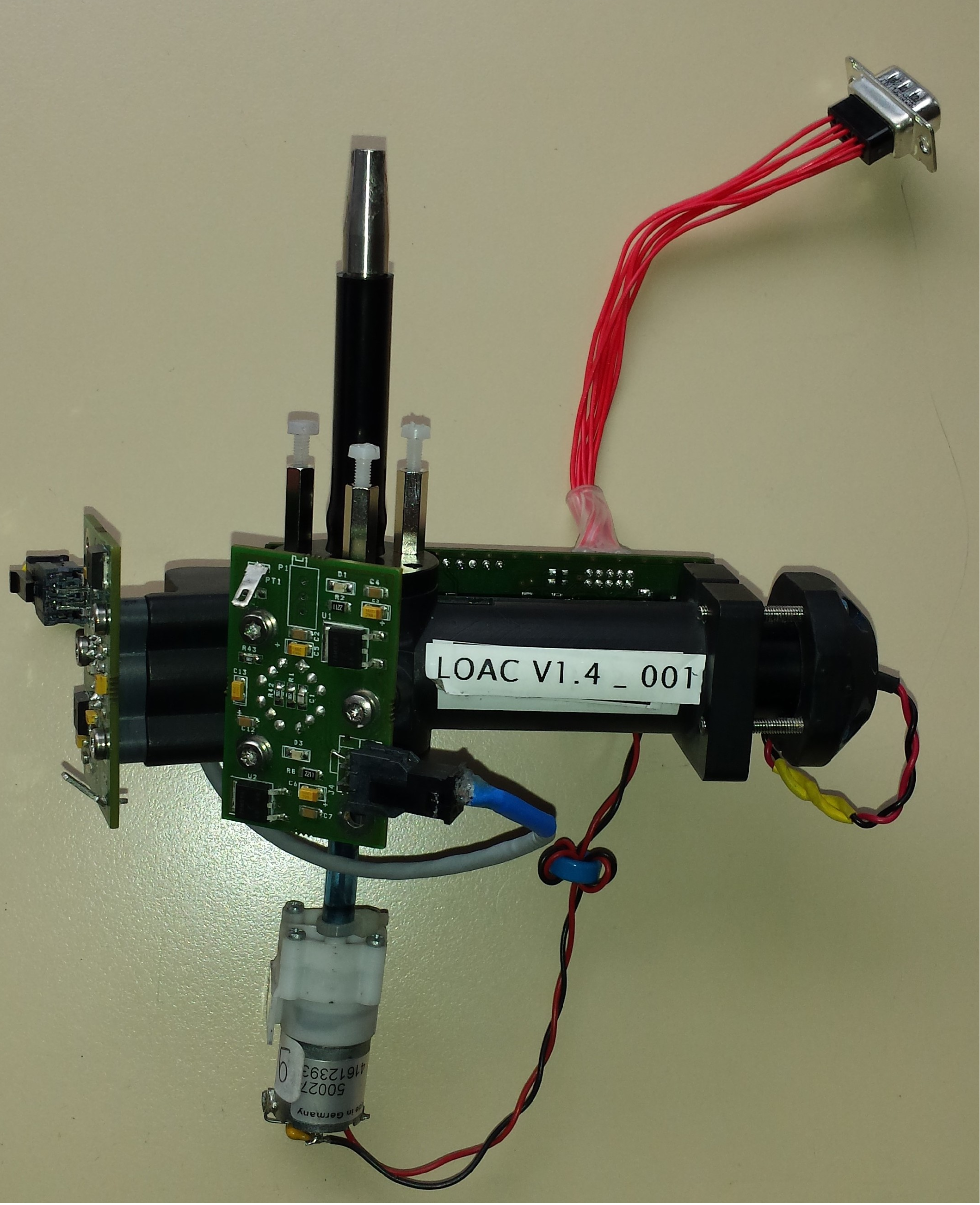}
\caption{The LOAC instrument used at present for short and long duration balloon flights. This version performs measurements at two scattering angles, while more angles are expected for the space version LONSCAPE}
\label{fig:LOAC}
\end{figure} 

\paragraph{Ortho-Para Instrument} 

Vertical mixing in giant planet tropospheres carrying significant heat from the deeper atmospheres to upper levels where it can be radiated to space is modulated by the atmospheric stability and can be dramatically changed by the condensation and evaporation of CH$_4$, H$_2$S, NH$_3$, and H$_2$O. Thermal profiles and stabilities in the colder outer solar system can be further affected by the atmospheric hydrogen para-fraction \cite{Sm95}. Hydrogen molecules come in two types -- with proton spins aligned (ortho-hydrogen) or opposite (para-hydrogen), each with significantly different thermodynamic properties at low temperatures. To interpret the thermal profile and stability, density structure, aerosol layering, net fluxes and vertical motions of giant planet atmospheres, the hydrogen para-fraction must be known, with increasing importance for the colder ice giants. The ortho- to para-hydrogen ratio can be measured by exploiting the thermodynamic differences between these two forms of hydrogen, which affects the speed of sound. Assuming atmospheric temperature and mean molecular weight are known, the ortho- to para-hydrogen ratio can be found from speed of sound measurements using a pair of ultrasonic capacitive transducers and sophisticated signal processing techniques. Acoustic travel times can be measured to $\sim$10 ns for travel times in the 0.5ms range (one part in 5e-4) using a high TRL, compact, energy-efficient and low data volume ultrasonic anemometer originally developed for Mars \cite{Ba16}.

\paragraph{Net Energy Flux Radiometer} Giant planet meteorology re- gimes depend on internal heat flux levels. Downwelling solar insolation and upwelling thermal energy from the planetary interior can have altitude and location dependent variations. Such radiative-energy differences cause atmospheric heating and cooling, and result in buoyancy differences that are the primary driving force for giant planets atmospheric motions. Three notable Net Flux Radiometer (NFR) instruments have flown in the past namely, the Large probe Infrared Radiometer (LIR) \cite{Bo80} on the Venus Probe, the NFR on the Galileo Probe \cite{Sr98}, and the DISR on the Huygens Probe \cite{To02}  for {\it in situ} measurements within Venus, Jupiter and Titan's atmospheres, respectively. All instruments were designed to measure the net radiative flux and upward radiation flux within their respective atmospheres as the probe descended by parachute. A future Net Flux Radiometer could build on the lessons learned from the Galileo probe NFR experiment and is designed to determine the net radiation flux within all giant planets atmospheres. The nominal measurement regime for the NFR extends from $\sim$0.1 bar to at least 10 bars. These measurements will help us to define sources and sinks of planetary radiation, regions of solar energy deposition, and provide constraints on atmospheric composition and cloud layers. The primary objective of the NFR is to measure upward and downward radiative fluxes to determine the radiative heating (cooling) component of the atmospheric energy budget, determine total atmospheric opacity, identify the location of cloud layers and opacities, and identify key atmospheric absorbers such as methane, ammonia, and water vapor. The NFR can measure upward and downward flux densities in multiple spectral channels.

\section{International Collaboration}

Only ESA/Europe and NASA/USA collaborations are considered here. However collaborations with other international partners may be envisaged. For several reasons, the participation of and contributions from NASA are essential for an ESA-led entry probe. NASA has proven its ability to send spacecrafts beyond 5 AU thanks to the use of radioisotope power systems. Although solar panel technologies likely enable the sending of spacecraft up to the distance of Saturn over the next decade, radioisotope power systems are required to reach the heliocentric distances of the Ice Giants. Also, because of their (relatively) small sizes, probes are ideal companion spacecraft to be included in ambitious missions similar to Cassini-Huygens or Galileo. An ESA giant planet probe mission could begin its flight phase as an element of a NASA Saturn, Uranus or Neptune mission (likely a NASA Flagship or New Frontiers mission). The launch would place both the NASA spacecraft, which functions also as the probe's CRSC, and the probe on a transfer trajectory to the giant planets. One of the key probe technologies for an entry probe that is critical for European industry is the heat shield material. If the European TPS is too heavy, then alternative material such as the NASA HEEET could be utilized in the context of a partnership between ESA and NASA.

\section{Education and Public Outreach (EPO)} The interest of the public in the giant planets continues to be significant, with much of the credit for the high interest in Saturn and Jupiter, due to the extraordinary success of the Cassini–Huygens mission and the currently ongoing Juno mission. Images from the Saturnian system and Jupiter are regularly featured as the NASA ``Astronomy Picture of the Day'', and continue to attract the interest of the international media. The interest and excitement of students and the general public can only be amplified by a return to Saturn or an unprecedented mission toward Uranus and/or Neptune. An entry probe mission will hold appeal for students at all levels. Education and Public Outreach activities will be an important part of the mission planning. An EPO team will be created to develop programs and activities for the general public and students of all ages. Additionally, results and interpretation of the science will be widely distributed to the public through internet sites, leaflets, public lectures, TV and radio programmes, numerical supports, museum and planetarium exhibitions, and in popular science magazines and in newspapers.

\section{Summary and Perspectives}

The next great planetary exploration mission may well be a flagship mission to Saturn, or one of the ice giant planets. This could be possibly a mission to Uranus with its unique obliquity and correspondingly extreme planetary seasons, its unusual dearth of cloud features and radiated internal energy, a tenuous ring system and multitude of small moons, or to the Neptune system, with its enormous winds, system of ring arcs, sporadic atmospheric features, and large retrograde moon Triton, likely a captured dwarf planet. The ice giant planets represent the last unexplored class of planets in the solar system, yet the most frequently observed type of exoplanets. Extended studies of Saturn, or one or both ice giants, including {\it in situ} measurements with an entry probe, are necessary to further constrain models of solar system formation and chemical, thermal, and dynamical evolution, the atmospheric formation, evolution, and processes, and to provide additional ground-truth for improved understanding of extrasolar planetary systems. The giant planets, gas and ice giants together, additionally offer a laboratory for studying the dynamics, chemistry, and processes of the terrestrial planets, including Earth's atmosphere. Only {\it in situ} exploration by a descent probe (or probes) can unlock the secrets of the deep, well-mixed atmospheres where pristine materials from the epoch of solar system formation can be found. Particularly important are the noble gases, undetectable by any means other than direct sampling, that carry many of the secrets of giant planet origin and evolution. Both absolute as well as relative abundances of the noble gases are needed to understand the properties of the interplanetary medium at the location and epoch of solar system formation, the delivery of heavy elements to the giant planet atmospheres, and to help decipher evidence of possible giant planet migration. A key result from a Saturn, Uranus or Neptune entry probe would be the indication as to whether the enhancement of the heavier noble gases found by the Galileo probe at Jupiter (and hopefully confirmed by a future Saturn probe) is a feature common to all the giant planets, or is limited only to the largest gas giant. This could have broad implications for the properties of known exoplanets of both giant and ice types, specially in planetary systems sharing both types of exoplanets.

The primary goal of a giant planet entry probe mission is to measure the well-mixed abundances of the noble gases He, Ne, Ar, Kr, Xe and their isotopes, the heavier elements C, N, S, and P, key isotope ratios $^{15}$N/$^{14}$N, $^{13}$C/$^{12}$C, $^{17}$O/$^{16}$O and $^{18}$O/$^{16}$O, and D/H, and disequilibrium species CO and PH$_3$, which act as tracers of internal processes, and can be achieved by a probe reaching 10 bars. In addition to measurements of the noble gases, chemical, and isotopic abundances in the atmosphere, a probe would measure many of the chemical and dynamical processes within the upper atmosphere, providing an improved context for understanding the chemistries, processes, origin, and evolution of all the atmospheres in the solar system. Moreover, the choice of an ice giant (Uranus or Neptune) entry probe would allow understanding the formation conditions of the entire family of all giant planets, and to provide ground-truth measurement to improve understanding of extrasolar planets. A descent probe would sample atmospheric regions far below those accessible to remote sensing, well into the cloud forming regions of the troposphere to depths where many cosmogenically important and abundant species are expected to be well-mixed. Along the descent, the probe would provide direct tracking of the planet's atmospheric dynamics including zonal winds, waves, convection and turbulence, measurements of the thermal profile and stability of the atmosphere, and the location, density, and composition of the upper cloud layers. Results obtained from a giant planet entry probe, and more importantly from an ice giant probe, are necessary to improve our understanding of the processes by which all the giants formed, including the composition and properties of the local solar nebula at the time and location of ice giant formation. By extending the legacy of the Galileo probe mission, Saturn, Uranus and/or Neptune probe(s) will further discriminate competing theories addressing the formation, and chemical, dynamical, and thermal evolution of the giant planets, the entire solar system including Earth and the other terrestrial planets, and the formation of other planetary systems.

\phantomsection
\section*{Acknowledgments} 

\addcontentsline{toc}{section}{Acknowledgments} 

O.M. acknowledges support from CNES. L.N.F. was supported by a Royal Society Research Fellowship. R.H. and A.S.L were  supported by the Spanish MINECO project AYA2015-65041-P (MINECO/FEDER, UE) and Grupos Gobierno Vasco IT-1366-19 from Gobierno Vasco. 
\clearpage

\phantomsection


\begin{thebibliography}{}

\bibitem{Go05} Gomes, R., Levison, H.~F., Tsiganis, K., et al.\ 2005, Nature, 435, 466
\bibitem{Ch01} Chambers, J.~E., \& Wetherill, G.~W.\ 2001, Meteoritics and Planetary Science, 36, 381
\bibitem{Ow99} Owen, T., Mahaffy, P., Niemann, H.~B., et al.\ 1999, Nature, 402, 269
\bibitem{Mo18} Mousis, O., Atkinson, D.~H., Cavali{\'e}, T., et al.\ 2018, Planetary and Space Science, 155, 12
\bibitem{vo98} von Zahn, U., Hunten, D.~M., \& Lehmacher, G.\ 1998, Journal of Geophysical Research, 103, 22815
\bibitem{Ow01} Owen, T., Mahaffy, P.~R., Niemann, H.~B., et al.\ 2001, The Astrophysical Journal Letters, 553, L77
\bibitem{Or98} Orton, G.~S., Fisher, B.~M., Baines, K.~H., et al.\ 1998, Journal of Geophysical Research, 103, 22791
\bibitem{Wo04} Wong, M.~H., Mahaffy, P.~R., Atreya, S.~K., et al.\ 2004, Icarus, 171, 153
\bibitem{Ho19} Hofstadter, M., Simon, A., Atreya, S., et al.\ 2019, Planetary and Space Science, in press
\bibitem{Ba18z} Banfield, D., Simon, A., Danner, R., et al.\ 2018, 2018 IEEE Aerospace Conference, Big Sky, MT, 2018, pp. 1-15, 10.1109/AERO.2018.8396829
\bibitem{Po96} Pollack, J.~B., Hubickyj, O., Bodenheimer, P., et al.\ 1996, Icarus, 124, 62
\bibitem{Al05a} Alibert, Y., Mousis, O., \& Benz, W.\ 2005, The Astrophysical Journal Letters, 622, L145
\bibitem{Al05b} Alibert, Y., Mousis, O., Mordasini, C., et al.\ 2005, The Astrophysical Journal Letters, 626, L57
\bibitem{Do10} Dodson-Robinson, S.~E., \& Bodenheimer, P.\ 2010, Icarus, 207, 491
\bibitem{He14a} Helled, R., Bodenheimer, P., Podolak, M., et al.\ 2014, Protostars and Planets VI, 643
\bibitem{Mi78} Mizuno, H., Nakazawa, K., \& Hayashi, C.\ 1978, Progress of Theoretical Physics, 60, 699
\bibitem{Li86} Lin, D.~N.~C., \& Papaloizou, J.\ 1986, The Astrophysical Journal, 307, 395
\bibitem{wa97} Ward, W.~R.\ 1997, Icarus, 126, 261
\bibitem{Id04} Ida, S., \& Lin, D.~N.~C.\ 2004, The Astrophysical Journal, 616, 567
\bibitem{Mor12} Mordasini, C., Alibert, Y., Klahr, H., et al.\ 2012, Astronomy \& Astrophysics, 547, A111
\bibitem{Bo97} Boss, A.~P.\ 1997, Science, 276, 1836
\bibitem{Bo01} Boss, A.~P.\ 2001, The Astrophysical Journal, 563, 367
\bibitem{Va18} Vazan, A., Helled, R., \& Guillot, T.\ 2018, Astronomy \& Astrophysics, 610, L14
\bibitem{Wa17} Wahl, S.~M., Hubbard, W.~B., Militzer, B., et al.\ 2017, Geophysical research Letters, 44, 4649
\bibitem{Ne17} Nettelmann, N.\ 2017, Astronomy \& Astrophysics, 606, A139
\bibitem{He13} Helled, R., \& Guillot, T.\ 2013, The Astrophysical Journal, 767, 113
\bibitem{Wi12a} Wilson, H.~F., \& Militzer, B.\ 2012, Physical Review Letters, 108, 111101
\bibitem{Wi12b} Wilson, H.~F., \& Militzer, B.\ 2012, The Astrophysical Journal, 745, 54
\bibitem{Le12} Leconte, J., \& Chabrier, G.\ 2012, Astronomy \& Astrophysics, 540, A20
\bibitem{Le13} Leconte, J., \& Chabrier, G.\ 2013, Nature Geoscience, 6, 347
\bibitem{Sa04} Saumon, D., \& Guillot, T.\ 2004, The Astrophysical Journal, 609, 1170
\bibitem{Fo10} Fortney, J.~J., \& Nettelmann, N.\ 2010, Space Science Reviews, 152, 423
\bibitem{Ne13} Nettelmann, N., Helled, R., Fortney, J.~J., et al.\ 2013, Planetary and Space Science, 77, 143
\bibitem{Ga01} Gautier, D., Hersant, F., Mousis, O., et al.\ 2001, The Astrophysical Journal Letters, 550, L227
\bibitem{He01} Hersant, F., Gautier, D., \& Hur{\'e}, J.-M.\ 2001, The Astrophysical Journal, 554, 391
\bibitem{yo98} Young, R.~E.\ 1998, Journal of Geophysical Research, 103, 22775
\bibitem{Fo98} Folkner, W.~M., Woo, R., \& Nandi, S.\ 1998, Journal of Geophysical Research, 103, 22847
\bibitem{Ra98} Ragent, B., Colburn, D.~S., Rages, K.~A., et al.\ 1998, Journal of Geophysical Research, 103, 22891
\bibitem{At98} Atkinson, D.~H., Pollack, J.~B., \& Seiff, A.\ 1998, Journal of Geophysical Research, 103, 22911
\bibitem{Sr98} Sromovsky, L.~A., Collard, A.~D., Fry, P.~M., et al.\ 1998, Journal of Geophysical Research, 103, 2929
\bibitem{Ni98} Niemann, H.~B., Atreya, S.~K., Carignan, G.~R., et al.\ 1998, Journal of Geophysical Research, 103, 22831
\bibitem{At99} Atreya, S.~K., Wong, M.~H., Owen, T.~C., et al.\ 1999, Planetary and Space Science, 47, 1243
\bibitem{Wi10} Wilson, H.~F., \& Militzer, B.\ 2010, Physical Review Letters, 104, 121101
\bibitem{Gu05} Guillot, T.\ 2005, Annual Review of Earth and Planetary Sciences, 33, 493
\bibitem{He06} Helled, R., Podolak, M., \& Kovetz, A.\ 2006, Icarus, 185, 64
\bibitem{At03} Atreya, S.~K., Mahaffy, P.~R., Niemann, H.~B., et al.\ 2003, Planetary and Space Science, 51, 105
\bibitem{Ma07} Matousek, S.\ 2007, Acta Astronautica, 61, 932
\bibitem{He14b} Helled, R., \& Lunine, J.\ 2014, Monthly Notices of the Royal Astronomical Society, 441, 2273
\bibitem{Co84} Conrath, B.~J., Gautier, D., Hanel, R.~A., et al.\ 1984, The Astrophysical Journal, 282, 807
\bibitem{Co00} Conrath, B.~J., \& Gautier, D.\ 2000, Icarus, 144, 124
\bibitem{Ac16} Achterberg, R.~K., Schinder, P.~J., \& Flasar, F.~M.\ 2016, AAS/Division for Planetary Sciences Meeting Abstracts \#48, 508.01
\bibitem{Cou84} Courtin, R., Gautier, D., Marten, A., et al.\ 1984, The Astrophysical Journal, 287, 899
\bibitem{Da96} Davis, G.~R., Griffin, M.~J., Naylor, D.~A., et al.\ 1996, Astronomy \& Astrophysics, 315, L393
\bibitem{Fl07} Fletcher, L.~N., Irwin, P.~G.~J., Teanby, N.~A., et al.\ 2007, Icarus, 188, 72
\bibitem{Fl09a} Fletcher, L.~N., Orton, G.~S., Teanby, N.~A., et al.\ 2009, Icarus, 202, 543
\bibitem{Fl09b} Fletcher, L.~N., Orton, G.~S., Teanby, N.~A., et al.\ 2009, Icarus, 199, 351
\bibitem{Mo14a} Mousis, O., Fletcher, L.~N., Lebreton, J.-P., et al.\ 2014, Planetary and Space Science, 104, 29
\bibitem{Mo16} Mousis, O., Atkinson, D.~H., Spilker, T., et al.\ 2016, Planetary and Space Science, 130, 80
\bibitem{At16} Atkinson, D.~H., Simon, A.~A., Banfield, D., et al.\ 2016, AAS/Division for Planetary Sciences Meeting Abstracts \#48, 123.29
\bibitem{Co87} Conrath, B., Gautier, D., Hanel, R., et al.\ 1987, Journal of Geophysical Research, 92, 15003
\bibitem{Bu03} Burgdorf, M., Orton, G.~S., Davis, G.~R., et al.\ 2003, Icarus, 164, 244
\bibitem{Ma00} Mahaffy, P.~R., Niemann, H.~B., Alpert, A., et al.\ 2000, Journal of Geophysical Research, 105, 15061
\bibitem{Li87} Lindal, G.~F., Lyons, J.~R., Sweetnam, D.~N., et al.\ 1987, Journal of Geophysical Research, 92, 14987
\bibitem{ba95} Baines, K.~H., Mickelson, M.~E., Larson, L.~E., et al.\ 1995, Icarus, 114, 328
\bibitem{Ka09} Karkoschka, E., \& Tomasko, M.\ 2009, Icarus, 202, 287
\bibitem{Sr14} Sromovsky, L.~A., Karkoschka, E., Fry, P.~M., et al.\ 2014, Icarus, 238, 137
\bibitem{Li90} Lindal, G.~F., Lyons, J.~R., Sweetnam, D.~N., et al.\ 1990, Geophysical Research Letters, 17, 1733
\bibitem{ka11} Karkoschka, E., \& Tomasko, M.~G.\ 2011, Icarus, 211, 780
\bibitem{Fl11} Fletcher, L.~N., Baines, K.~H., Momary, T.~W., et al.\ 2011, Icarus, 214, 510
\bibitem{Ir18} Irwin, P.~G.~J., Toledo, D., Garland, R., et al.\ 2018, Nature Astronomy, 2, 420
\bibitem{Ir19} Irwin, P.~G.~J., Toledo, D., Garland, R., et al.\ 2019, Icarus, 321, 550
\bibitem{Lo09} Lodders, K., Palme, H., \& Gail, H.-P.\ 2009, Landolt B\"ornstein, 4B, 712
\bibitem{Fe13} Feuchtgruber, H., Lellouch, E., Orton, G., et al.\ 2013, Astronomy \& Astrophysics, 551, A126
\bibitem{Al14} Ali-Dib, M., Mousis, O., Petit, J.-M., et al.\ 2014, The Astrophysical Journal, 793, 9
\bibitem{Ge98} Geiss, J., \& Gloeckler, G.\ 1998, Space Science Reviews, 84, 239
\bibitem{Ma11} Marty, B., Chaussidon, M., Wiens, R.~C., et al.\ 2011, Science, 332, 1533
\bibitem{Fo00} Fouchet, T., Lellouch, E., B{\'e}zard, B., et al.\ 2000, Icarus, 143, 223
\bibitem{Fl14} Fletcher, L.~N., Greathouse, T.~K., Orton, G.~S., et al.\ 2014, Icarus, 238, 170
\bibitem{Mo14b} Mousis, O., Lunine, J.~I., Fletcher, L.~N., et al.\ 2014, The Astrophysical Journal, 796, L28
\bibitem{Ru15} Rubin, M., Altwegg, K., Balsiger, H., et al.\ 2015, Science, 348, 232
\bibitem{Le15} Le Roy, L., Altwegg, K., Balsiger, H., et al.\ 2015, Astronomy \& Astrophysics, 583, A1
\bibitem{Bo04} Bockel{\'e}e-Morvan, D., Crovisier, J., Mumma, M.~J., et al.\ 2004, Comets II, 391
\bibitem{Ro14} Rousselot, P., Pirali, O., Jehin, E., et al.\ 2014, The Astrophysical Journal Letters, 780, L17
\bibitem{Ma09} Manfroid, J., Jehin, E., Hutsem{\'e}kers, D., et al.\ 2009, Astronomy \& Astrophysics, 503, 613
\bibitem{Ma98} Mahaffy, P.~R., Donahue, T.~M., Atreya, S.~K., et al.\ 1998, Space Science Reviews, 84, 251
\bibitem{Le01} Lellouch, E., B{\'e}zard, B., Fouchet, T., et al.\ 2001, Astronomy \& Astrophysics, 370, 610
\bibitem{Do18} Dobrijevic, M., \& Loison, J.~C.\ 2018, Icarus, 307, 371
\bibitem{As09} Asplund, M., Grevesse, N., Sauval, A.~J., et al.\ 2009, Annual Reviews of Astronomy \& Astrophysics, 47, 481
\bibitem{Bo12} Bockel{\'e}e-Morvan, D., Biver, N., Swinyard, B., et al.\ 2012, Astronomy \& Astrophysics, 544, L15
\bibitem{Co11} Courtin, R., Swinyard, B.~M., Moreno, R., et al.\ 2011, Astronomy \& Astrophysics, 536, L2
\bibitem{Lo17} Loison, J.~C., Dobrijevic, M., Hickson, K.~M., et al.\ 2017, Icarus, 291, 17
\bibitem{Se16} Serigano, J., Nixon, C.~A., Cordiner, M.~A., et al.\ 2016, The Astrophysical Journal Letters, 821, L8
\bibitem{No95} Noll, K.~S., Geballe, T.~R., \& Knacke, R.~F.\ 1995, The Astrophysical Journal Letters, 453, L49
\bibitem{Bo02} Boss, A.~P., Wetherill, G.~W., \& Haghighipour, N.\ 2002, Icarus, 156, 291
\bibitem{Bar07} Bar-Nun, A., Notesco, G., \& Owen, T.\ 2007, Icarus, 190, 655
\bibitem{Mo10} Mousis, O., Lunine, J.~I., Picaud, S., et al.\ 2010, Faraday Discussions, 147, 509
\bibitem{Mo12} Mousis, O., Lunine, J.~I., Madhusudhan, N., et al.\ 2012, The Astrophysical Journal Letters, 751, L7
\bibitem{Gu06} Guillot, T., \& Hueso, R.\ 2006, Monthly Notices of the Royal Astronomical Society, 367, L47
\bibitem{St88} Stevenson, D.~J., \& Lunine, J.~I.\ 1988, Icarus, 75, 146
\bibitem{Cy98} Cyr, K.~E., Sears, W.~D., \& Lunine, J.~I.\ 1998, Icarus, 135, 537
\bibitem{Gu95} Guillot, T.\ 1995, Science, 269, 1697
\bibitem{Vi05} Visscher, C., \& Fegley, B.\ 2005, The Astrophysical Journal, 623, 1221
\bibitem{Ca17} Cavali{\'e}, T., Venot, O., Selsis, F., et al.\ 2017, Icarus, 291, 1

\bibitem{Sanchez-Lavega2019book} S\'anchez-Lavega, A. et al.\ 2019, in Zonal Jets: Phenomenology, Genesis and Physics, Cambrige University PRess, eds: B. Galperin and P.~L. Read.
\bibitem{Vasavada2005} Vasavada, A. R., Showman, A. P.\ 2005, Reports on Progress in Physics, 68, 1935
\bibitem{Kaspi2018} Kaspi, Y., Galanti, E., Hubbard, W. B. et al.\ 2018, Nature, 555, 223
\bibitem{Guillot2018} Guillot, T., Miguel, Y., Millitzer, B. et al.\ 2018, Nature, 555, 227
\bibitem{Galanti2019} Galanti, E., Kaspi, Y., Miguel, Y. et al.\ 2019, Geophysical Research Letters, 46, 616
\bibitem{Kaspi2013} Kaspi, Y., Showman, A. P., Hubbard, W. B. et al.\ 2013, Nature, 497, 344
\bibitem{Atkinson1996} Atkinson, D. H., Pollack, J. B., Seiff, A.\ 1996, Science, 272, 842
\bibitem{Atkinson1997} Atkinson, D. H., Ingersoll, A. P., Seiff, A.\ 1997, Nature, 388, 649
\bibitem{Garcia-Melendo2011} Garcia-Melendo, E., P\'erez-Hoyos, S., S\'anchez-Lavega, A. et al.\ Icarus, 215, 62
\bibitem{Tollefson2018} Tollefson, J., de Pater, I., Marcus, P. S. et al.\ 2018, Icarus, 311, 317
\bibitem{Molter2019} Molter, E., de Pater, I., Luszcz-Cook, S. et al.\ 2019, Icarus, 321, 324
\bibitem{Sun1991} Sun, Z.~P., Schubert, G., Stoker, C. R.\ 1991, Icarus, 91, 154
\bibitem{Lian2010} Lian Y. and A. P. Showman\ 2010, Icarus, 207, 373
\bibitem{Fletcher2007} Fletcher, L. N., Irwin, P.G.J., Teanby, N. A., et al.\ 2007, Icarus, 189, 457
\bibitem{Orton2014} Orton, G. S. Fletcher, L. N., Moses, J. I. et al.\ 2014, Icarus, 243, 494
\bibitem{Fletcher2014} Fletcher, L. N., de Pater, I., Orton, G. S. et al.\ 2014, Icarus, 231, 146
\bibitem{Lindal1987} Lindal, G. F., Lyons, J.R., Sweetnam, D. N. et al.\ 1987, Journal of Geophys. Res. 92, 14987 
\bibitem{Lindal1992} Lindal, G. F.\ 1992, The Astronomical Journal, 103, 967
\bibitem{Leconte2017} Leconte, K., Selsis, F., Hersant, F. et al.\ 2017, Astronomy \& Astrophysics, 598, A98
\bibitem{Friedson2017} Friedson, A.~J., \& Gonzales, E.~J.\ 2017, Icarus, 297, 160
\bibitem{Herbert1987} Herbert, F., Sandel, B. R., Yelle, R. V. et al.\ 1987, Journal of Geophys. Res., 92, 15093
\bibitem{Li2018} Li, C., Le, T., Zhang X., et al.\ 2018, Journal of Quantitative Spectroscopy and Radiative Transfer, 217, 353.
\bibitem{Banfield1998} Banfield, D., Gierasch, P. J., Bell, M. et al.\  1998, Icarus, 135, 230
\bibitem{West2004} West, R. A., Baines, K. H., Friedson, J.A. et al.\ 2004, in Jupiter the Planet, Satellites and Magnetosphere, Cambridge University Press, eds: Bagenal, F., Dowling, T. E., McKinnon, W. B.
\bibitem{Perez-Hoyos2012} P\'erez-Hoyos, S., Sanz-Requena, J.F., Barrado-Izagirre, N. et al.\ 2012, Icarus, 217, 256 
\bibitem{West2009} West, R. A., Baines, K. H., Karkoschka, E. et al.\ 2009, in Saturn from Cassini-Huygens, Cambridge University Press, eds: Dougherty, M. K., Esposito, L. W., Krimigis, S.M.
\bibitem{Fletcher2011} Fletcher, L. N., Baines, K. H., Momary, T. W. et al.\ 2011, Icarus, 214, 510
\bibitem{Perez-Hoyos2016} P\'erez-Hoyos, S., S\'anchez-Lavega, A., Irwin, P. G. J. et al.\ 2016, Icarus, 277, 1
\bibitem{West1991} West, R. A., Baines, K. H., Pollack, J. B., \ 1991, in Uranus, University of Arizona Press, eds: Bergstralh, J.T., Miner, E. D. Matthews, M.S. 
\bibitem{Irwin2009} Irwin, P. G. J., Teanby, N. A., Davis, G. R.\ 2009, Icarus, 203, 287 
\bibitem{deKleer2015} de Kleer, K., Luszcz-Cook, S., de Pater, I. et al. \ 2015, Icarus, 256, 120 
\bibitem{Irwin2017} Irwin, P. G. J., Wong, M. H. Simon, A. A. et al.\ 2017, Icarus, 288, 99 
\bibitem{Hammel1989} Hammel, H. B., Baines, K. H., Bergstralh, J. T.,\ 1989, Icarus, 80, 416 
\bibitem{Baines1994} Baines, K. H., Hammel, H. B.,\ 1994, Icarus, 109, 20 
\bibitem{Irwin2016} Irwin, P. G. J., Fletcher, L. N., Tice, D. et al.\ 2016, Icarus, 271, 418 
\bibitem{Ingersoll2004} Ingersoll, A. P., Dowling, T. E., Gierasch, P. J.,\ 2004, in Jupiter the Planet, Satellites and Magnetosphere, Cambridge University Press, eds: Bagenal, F., Dowling, T. E., McKinnon, W. B.
\bibitem{Vasavada2006} Vasavada, A. R., H{\"o}rst, S. M., Kennedy, M. R., et al.\ 2006, Journal of Geophysical Research (Planets) 111, 5004
\bibitem{Smith1986} Smith, B. A., Soderblom, L. A., Beebe, R. et al.\ 1986, Science, 233, 43
\bibitem{Smith1989} Smith, B. A., Soderblom, L. A., Banfield, D. et al.\ 1989, Science, 246, 1422
\bibitem{Sromovsky2001} Sromovsky, L. A., Fry, P. M., Dowling, T. E.\ 2001, Icarus, 149, 459
\bibitem{Fry2012} Fry, P. M., Sromovsky, L. A., de Pater, I., et al.\ 2012, The Astronomical Journal, 143, 150
\bibitem{Fletcher2019} Fletcher, L. N., Kaspi, Y., Guillot, T. et al.\ 2019, arXiv:1907:01822
\bibitem{Weidenschilling1973} Weidenschilling, S. J., Lewis, J. S.\ 1973, Icarus, 20, 465
\bibitem{Atreya2005} Atreya, S. K., Wong, A. S.\ 2005, Space Sci. Rev., 116, 121
\bibitem{Irwin2018} Irwin, P. G. J., Toledo, D., Garland, R.\ 2018, Nature Astronomy, 2, 420
\bibitem{Irwin2019} Irwin, P. G. J., Toledo, D., Garland, R.\ 2019, Icarus, 321, 550
\bibitem{Showman1998} Showman, A. P., Ingersoll, A. P., Icarus, 132, 205
\bibitem{Fischer2019} Fischer, G., Pagaran, J. A., Zarka, P. et al.\ 2019, Astronomy \& Astrophysics, 621, A113
\bibitem{Sanchez-Lavega2018_book} S\'anchez-Lavega, A., Fischer, G., Fletcher, L. N. et al.\ 2018, in Saturn in the 21st Century, Cambridge University Press, eds: Baines, K. H., Flasar, F. M., Krupp, N. et al.
\bibitem{dePater2015} de Pater, I., Sromovsky, L. A., Fry, P. M. et al.\ 2015, Icarus, 252, 121
\bibitem{Irwin_Uranus2017} Irwin, P. G. J., Wong, M. H., Simon, A. A. et al.\ 2017, Icarus, 288, 99
\bibitem{Stoker1989} Stoker, C. R., Toon, O. B.\ 1989, Geophys. Res. Lett. 16, 929
\bibitem{Showman2018} Showman, A. P., Ingersoll, A. P., Achterberg, R. et al.\ 2018, in Saturn in the 21st Century, Cambridge University Press, eds: Baines, K. H., Flasar, F. M., Krupp, N. et al.
\bibitem{Wong2018} Wong, M. H., Tollefson, J., Hsu, A. I. et al.\ 2018, The Astronomical Journal, 155, 117
\bibitem{Hammel2009} Hammel, H. B., Sromovsky, L. A., Fry, P. M. et al.\ 2009 ,Icarus, 201, 257 
\bibitem{Sromovsky2015} Sromovsky, L.~A., de Pater, I., Fry, P. M. et al.\ 2015, Icarus, 258, 192
\bibitem{Orton1998} Orton, G. S., Fisher, B. M., Baines, K. H. et al.\ 1998, Journal of Geophysical Research, 103, 22791
\bibitem{Taylor2004} Taylor, F. W., Atreya, S. K., Encrenaz, Th. et al.\ 2004, in Jupiter the Planet, Satellites and Magnetosphere, Cambridge University Press, eds: Bagenal, F., Dowling, T. E., McKinnon, W. B.
\bibitem{Fouchet2009} Fouchet, T., Moses, J. I., Conrath, B. J. et al.\ 2009, in Saturn from Cassini-Huygens, Cambridge University Press, eds: Dougherty, M. K., Esposito, L. W., Krimigis, S.M.
\bibitem{Moses2018} Moses, K. I. Fletcher, L. N., Greathouse, T. K. et al.\ 2018, Icarus 307, 124

\bibitem{Vl16} Venkatapathy, E., Ellerby, D., Gage, P. \ 2019. In: Workshop on In Situ Exploration of Ice Giants, Marseille, France
\bibitem{Mi17} Milos, F.~S., Chen, Y.-K., \& Mahzari, M.\ 2017, Journal of Spacecraft and Rockets, In: 47th AIAA Thermophysics Conference, AIAA AVIATION Forum, https://doi.org/10.2514/6.2017-3353 
\bibitem{Fu02} Fulchignoni, M., Ferri, F., Angrilli, F., et al.\ 2002, Space Science Reviews, 104, 395
\bibitem{Se92} Seiff, A., \& Knight, T.~C.~D.\ 1992, Space Science Reviews, 60, 203
\bibitem{Se80} Seiff, A., Juergens, D.~W., \& Lepetich, J.~E.\ 1980, IEEE Transactions on Geoscience and Remote Sensing, 18, 105
\bibitem{Ba07} Balsiger, H., Altwegg, K., Bochsler, P., et al.\ 2007, Space Science Reviews, 128, 745
\bibitem{Du10} Durry, G., Li, J.~S., Vinogradov, I., et al.\ 2010, Applied Physics B: Lasers and Optics, 99, 339
\bibitem{We11} Webster, C.~R., \& Mahaffy, P.~R.\ 2011, Planetary and Space Science, 59, 271
\bibitem{We13} Webster, C.~R., Mahaffy, P.~R., Flesch, G.~J., et al.\ 2013, Science, 341, 260
\bibitem{vo92} von Zahn, U., \& Hunten, D.~M.\ 1992, Space Science Reviews, 60, 263
\bibitem{At97} Atkinson, D.~H., Ingersoll, A.~P., \& Seiff, A.\ 1997, Nature, 388, 649
\bibitem{Ga97} Gayet, J.~F., Cr{\'e}pel, O., Fournol, J.~F., et al.\ 1997, Annales Geophysicae, 15, 451
\bibitem{Lu14} Lurton, T., Renard, J.-B., Vignelles, D., et al.\ 2014, Atmospheric Measurement Techniques, 7, 931
\bibitem{Re16b} Renard, J.-B., Dulac, F., Berthet, G., et al.\ 2016, Atmospheric Measurement Techniques, 9, 3673
\bibitem{Re16a} Renard, J.-B., Dulac, F., Berthet, G., et al.\ 2016, Atmospheric Measurement Techniques, 9, 1721
\bibitem{Re02} Renard, J.-B., Berthet, G., Robert, C., et al.\ 2002, Applied Optics, 41, 7540
\bibitem{Vo06} Volten, H., Mu{\~n}oz, O., Hovenier, J.~W., et al.\ 2006, J. Quant. Spectrosc. Radiat. Transf., 100, 437
\bibitem{Sm95} Smith, M.~D., \& Gierasch, P.~J.\ 1995, Icarus, 116, 159
\bibitem{Ba16} Banfield, D., Schindel, D.~W., Tarr, S., et al.\ 2016, Acoustical Society of America Journal, 140, 1420
\bibitem{Bo80} Boese, R.~W., Twarowski, R.~J., Gilland, J., et al.\ 1980, IEEE Transactions on Geoscience and Remote Sensing, 18, 97
\bibitem{To02} Tomasko, M.~G., Buchhauser, D., Bushroe, M., et al.\ 2002, Space Science Reviews, 104, 469


\end{thebibliography}

\clearpage
\section*{Core Proposing Team} 

\paragraph{Olivier Mousis} Aix Marseille Univ, CNRS, CNES, LAM, Marseille, France

\paragraph{David H. Atkinson} Jet Propulsion Laboratory, California Institute of Technology, 4800 Oak Grove Dr., Pasadena, CA, 91109, USA

\paragraph{Richard Ambrosi}Department of Physics and Astronomy, University of Leicester, Leicester, UK

\paragraph{Sushil Atreya} Department of Atmospheric, Oceanic, and Space Sciences, University of Michigan, Ann Arbor, MI, 48109-2143, USA

\paragraph{Don Banfield} Cornell Center for Astrophysics and Planetary Science, Cornell University, 420 Space Sciences, Ithaca, NY 14853, USA

\paragraph{Stas Barabash} Swedish Institute of Space Physics, Box 812, 98128, Kiruna, Sweden

\paragraph{Michel Blanc} Institut de Recherche en Astrophysique and
 Plan\'etologie (IRAP), CNRS/Universit\'e Paul Sabatier, 31028 Toulouse, France

\paragraph{Thibault Cavali\'e} Laboratoire d'astrophysique de Bordeaux, University Bordeaux, CNRS, B18N, all\'ee Geoffroy Saint-Hilaire, 33615 Pessac, France

\paragraph{Athena Coustenis} LESIA, Observatoire de Paris, PSL Research University, CNRS, Sorbonne Universit\'es, UPMC Univ. Paris 06, Univ. Paris Diderot, Sorbonne Paris Cit\'e, 5 place Jules Janssen, 92195 Meudon, France

\paragraph{Magali Deleuil} Aix Marseille Univ, CNRS, CNES, LAM, Marseille, France

\paragraph{Georges Durry} Groupe de Spectrométrie Mol\'eculaire et Atmosphérique, UMR 7331, CNRS, Universit\'e de Reims, Champagne Ardenne, Campus Sciences Exactes et Naturelles, BP 1039, Reims 51687, France

\paragraph{Francesca Ferri} Universit\`a degli Studi di Padova, Centro di Ateneo di Studi e Attivit\`a Spaziali ``Giuseppe Colombo'' (CISAS), via Venezia 15, 35131 Padova, Italy

\paragraph{Leigh Fletcher} Department of Physics \& Astronomy, University of Leicester, University Road, Leicester, LE1 7RH, UK

\paragraph{Thierry Fouchet} LESIA, Observatoire de Paris, PSL Research University, CNRS, Sorbonne Universit\'es, UPMC Univ. Paris 06, Univ. Paris Diderot, Sorbonne Paris Cit\'e, 5 place Jules Janssen, 92195 Meudon, France

\paragraph{Tristan Guillot} Observatoire de la C\^ote d'Azur, Laboratoire Lagrange, BP 4229, 06304 Nice cedex 4, France

\paragraph{Paul Hartogh} Max-Planck-Institut f\"ur Sonnensystemforschung, Justus von Liebig Weg 3, 37077 G\"ottingen, Germany

\paragraph{Ricardo Hueso} Escuela de Ingenier\'ia de Bilbao, \\UPV/EHU, 48013 Bilbao, Spain

\paragraph{Mark Hofstadter} Jet Propulsion Laboratory, California Institute of Technology, 4800 Oak Grove Dr., Pasadena, CA, 91109, USA

\paragraph{Jean-Pierre Lebreton} CNRS-Universit{\'e} d'Orl{\'e}ans, 3a Avenue de la Recherche Scientifique, 45071 Orl{\'e}ans Cedex 2, France

\paragraph{Kathleen E. Mandt} Applied Physics Laboratory, Johns Hopkins University, 11100 Johns Hopkins Rd., Laurel, MD 20723, USA

\paragraph{Heike Rauer} Institut f\"ur Planetenforschung, DLR, Berlin; Zentrum f\"ur Astronomie und Astrophysik, TU Berlin, Berlin

\paragraph{Pascal Rannou} Groupe de Spectrométrie Mol\'eculaire et Atmosphérique, UMR 7331, CNRS, Universit\'e de Reims Champagne Ardenne, Campus Sciences Exactes et Naturelles, BP 1039, Reims 51687, France

\paragraph{Jean-Baptiste Renard} CNRS-Universit{\'e} d'Orl{\'e}ans, 3a Avenue de la Recherche Scientifique, 45071 Orl{\'e}ans Cedex 2, France

\paragraph{Agustin S\'anchez-Lavega} Escuela de Ingenier\'ia de Bilbao, UPV/EHU, 48013 Bilbao, Spain

\paragraph{Kunio Sayanagi} Department of Atmospheric and Planetary Sciences, Hampton University, 23 Tyler Street, Hampton, VA 23668, USA

\paragraph{Amy Simon} NASA Goddard Space flight Center, Greenbelt, MD, 20771, USA

\paragraph{Thomas Spilker} Solar System Science \& Exploration, Monrovia, USA

\paragraph{Ethiraj Venkatapathy} NASA Ames Research Center, Moffett field, CA, USA

\paragraph{J. Hunter Waite} Southwest Research Institute, San Antonio, TX, 78228, USA

\paragraph{Peter Wurz} Space Science \& Planetology, Physics Institute, University of Bern, Sidlerstrasse 5, 3012 Bern, Switzerland\\ \\ \\ 

\noindent Cover illustration by courtesy of Tibor Balint

\end{document}